\documentclass[12pt]{iopart}
\usepackage{graphicx}
\usepackage{iopams}
\eqnobysec

\begin{document}


\title[Low-energy expansion formula for periodic potentials]
{Low-energy expansion formula 
for one-dimensional Fokker-Planck and Schr\"odinger equations
with periodic potentials}

\author{Toru Miyazawa}

\address{Department of Physics, Gakushuin University, 
Tokyo 171-8588, Japan}
\ead{toru.miyazawa@gakushuin.ac.jp}
\begin{abstract}
We study the low-energy behavior of the Green function for one-dimensional Fokker-Planck and Schr\"odinger equations with periodic potentials. We derive a formula for the power series expansion of reflection coefficients in terms of the wave number, and apply it to the low-energy expansion of the Green function. 
\end{abstract}

\maketitle


\section{Introduction}
One-dimensional diffusion in a potential $V(x)$ is described by the Fokker-Planck equation
\begin{equation}
\label{1-1.1}
-\frac{\rmd^2}{\rmd x^2}\phi(x)+2\frac{\rmd}{\rmd x}[f(x)\phi(x)]=k^2\phi(x),
\end{equation}
where
\begin{equation}
\label{1-1.2}
f(x)=-\frac{1}{2}\frac{\rmd}{\rmd x}V(x).
\end{equation}
The Fokker-Planck equation (\ref{1-1.1})
is equivalent to the Schr\"odinger equation
\begin{equation}
\label{1-1.3}
-\frac{\rmd^2}{\rmd x^2}\psi(x)+V_{\rm S}(x)\psi(x)=k^2\psi(x),
\end{equation}
where $\psi$ and $V_{\rm S}$ are related to $\phi$ and $f$ in (\ref{1-1.1}) by
\begin{equation}
\label{1-1.4}
\phi(x)=\rme^{-V(x)/2}\psi(x), \qquad V_{\rm S}(x)=f^2(x)+f'(x).
\end{equation}

We assume that ${\rm Im}\,k \geq 0$. Let $G_{\rm S}(x,y;k)$ denote the Green function for the Schr\"odinger equation, satisfying
\begin{equation}
\label{1-1.5}
\left[\frac{\partial^2}{\partial x^2}-V_{\rm S}(x) +k^2 \right]G_{\rm S}(x,y;k)
=\delta(x-y)
\end{equation}
together with the boundary conditions $G_{\rm S}(x,y;k)\to 0$ as $\vert x-y \vert \to \infty$ for ${\rm Im}\,k>0$. 
For real $k$, we define $G_{\rm S}(x,y;k)\equiv \lim_{\epsilon \downarrow 0}G_{\rm S}(x,y;k+\rm i\epsilon)$. Without loss of generality we assume $x \geq y$.
The Green function for the Fokker-Planck equation is given by
\begin{equation}
\label{1-1.6}
G_{\rm F}(x,y;k) \equiv \rme^{-[V(x)-V(y)]/2}G_{\rm S}(x,y;k).
\end{equation}
This $G_{\rm F}(x,y;k)$ satisfies
\begin{equation}
\label{1-1.7}
\left[\frac{\partial^2}{\partial x^2}-2\frac{\partial}{\partial x}f(x) +k^2 \right]G_{\rm F}(x,y;k)
=\delta(x-y).
\end{equation}

In previous papers \cite{analysis,low}, we introduced a new method for systematically calculating the expansion of $G_{\rm S}(k)$ in powers of $k$.
In these papers, it was assumed that the potential $V(x)$ either steadily tends to a finite limit, or steadily diverges to $+\infty$ or $-\infty$, as $x \to \pm \infty$.
We wish to extend our method to be applicable to more general potentials, including the cases where $V(x)$ oscillates as $x \to \pm \infty$. 
In the present paper we study the simplest of such cases, namely, the cases where the potential is purely periodic.
We assume that the potential $V(x)$ is a real-valued periodic function with period~$L$, 
\begin{equation}
\label{1-1.8}
V(x+L)=V(x),
\end{equation}
and that $V(x)$ is piecewise continuously differentiable\footnote{
%
%
We allow $V(x)$ to have jump discontinuities. Although $V_{\rm S}(x)$ does not make sense when $V(x)$ is discontinuous, the Fokker-Planck equation is well defined even for such $V$. See \cite{risken} or footnote~1 of \cite{low}. 
When $V_{\rm S}$ does not make sense, $G_{\rm S}$ is defined by (\ref{1-1.7}) and (\ref{1-1.6}).
}.

The Fokker-Planck equation with a periodic potential has various applications such as Josephson junctions \cite{josephson} and superionic conductors \cite{superionic1,superionic2}, to name a few. And, needless to say, the Schr\"odinger equation with a periodic potential plays an essential role in solid state physics. 
Thus, the study of periodic systems is significant in itself.
Schr\"odinger operators with periodic potentials have been extensively investigated over the years, and much is known about their spectral properties [7--17].

To carry out the analysis of the Green function, we adopt a formalism that deals with reflection coefficients.
Reflection coefficients are of extreme importance in scattering theory, and their properties have been studied by many researchers, especially in connection with the inverse scattering method [12--16]. 
The formula derived in \cite{analysis} provides a new method for calculating the expansion of reflection coefficients up to an arbitrary order in $k$.
The aim of the present paper is to modify this formula and make it applicable to the periodic case.
The resulting formula serves as a new tool for the analysis of periodic systems, and enables us to deal with periodic and non-periodic potentials on the same basis.
Although we restrict ourselves to periodic potentials in this paper, this method can be further extended to more general potentials which show oscillatory behavior at infinity.

The necessary results from the previous papers will be briefly reviewed in section~3, after making some definitions in the next section.
In sections~4--8, we discuss how to adapt the expansion formula to the periodic case. 
Using the expansion of reflection coefficients, we shall derive the low-energy expansion of the Green function in sections~9 and 10. As is well known, the energy spectrum of a periodic system has a band structure. The expansion in powers of $k$ is an expansion from the bottom of the lowest band, and it is  effective for calculating the Green function in the lowest band.
An example is given in section~11.


\section{Preliminaries}
In this section we summarize the necessary definitions and notation.
We define the evolution matrix $U(x,x';k)$ as the $2 \times 2$ matrix satisfying the differential equation \cite{theory}
\begin{equation}
\label{1-2.1}
\frac{\partial}{\partial x}
U(x,x';k)=
\left(
\begin{array}{cc}
-\rmi k & f(x) \\
f(x) & \rmi k \\
\end{array}
\right)
U(x,x';k)
\end{equation}
with the initial condition
\begin{equation}
\label{1-2.2}
U(x',x';k)=
\left(
\begin{array}{cc}
1 & 0 \\
0 & 1 \\
\end{array}
\right).
\end{equation}
It is obvious from (\ref{1-2.1}) that the upper-right and lower-right elements of $U$ are obtained, respectively, from the lower-left and upper-left elements by changing the sign of $k$.
We write these elements as
\begin{equation}
\label{1-2.3}
U(x,x';k)\equiv
\left(
\begin{array}{cc}
\alpha(x,x';k) & \beta(x,x';-k) \\
\beta(x,x';k) & \alpha(x,x';-k) \\
\end{array}
\right).
\end{equation}
From (\ref{1-2.1}), we can easily show that $\alpha(x,x';k)+ \beta(x,x';k)$ and $\alpha(x,x';-k)+ \beta(x,x';-k)$, as functions of $x$, are solutions of the Schr\"odinger equation (\ref{1-1.3}). The matrix equation (\ref{1-2.1}) is thus equivalent to (\ref{1-1.1}) and (\ref{1-1.3}).

Since the matrix multiplying $U$ on the right-hand side of (\ref{1-2.1}) is traceless,
the determinant of $U$ is unity. Namely,
\begin{equation}
\label{1-2.4}
\alpha(x,x';k)\alpha(x,x';-k)-\beta(x,x';k)\beta(x,x';-k)=1.
\end{equation}
It is easy to see that the evolution matrix has the property
\begin{equation}
\label{1-2.5}
U(x_3,x_2;k)U(x_2,x_1;k)=U(x_3,x_1,;k).
\end{equation}
If the potential is a periodic function satisfying (\ref{1-1.8}), then $f(x+L)=f(x)$, and so
\begin{equation}
\label{1-2.6}
U(x+L,x'+L;k)=U(x,x';k).
\end{equation}

The transmission coefficient $\tau$, the right reflection coefficient $R_r$, and the left reflection coefficient $R_l$ are defined in terms of $\alpha$ and $\beta$ as
\refstepcounter{equation}
\label{1-2.7}
\addtocounter{equation}{-1}
\numparts
\begin{equation}
\tau(x,x';k)\equiv \frac{1}{\alpha(x,x';k)},
\end{equation}
\begin{equation}
\label{1-2.7b}
R_r(x,x';k)\equiv \frac{\beta(x,x';k)}{\alpha(x,x';k)},
\qquad
R_l(x,x';k)\equiv -\frac{\beta(x,x';-k)}{\alpha(x,x';k)}.
\end{equation}
\endnumparts
(See \cite{theory}. Similar construction of the reflection coefficients is used, for example, in \cite{r1}.)
Let us explain why these quantities are called the transmission and reflection coefficients.
We consider a potential which is identical to $V(x)$ inside the interval $(x_1,x_2)$ and defined to be constant outside this interval. Namely, we define
\begin{equation}
\fl
V_{x_1,x_2}(x)\equiv
\cases{
V(x_1) & $(x<x_1)$ \\
V(x) & $(x_1\leq x \leq x_2)$ \\
V(x_2) & $(x_2<x)$,
}
\qquad 
f_{x_1,x_2}(x)\equiv -\frac{1}{2}\frac{\rmd}{\rmd x} V_{x_1,x_2}(x).
\end{equation}
The Schr\"odinger equation corresponding to this Fokker-Planck potential $V_{x_1,x_2}$ is
\begin{equation}
\label{2-2.9}
-\frac{\rmd^2 \psi}{\rmd x^2}
+\left(f^2_{x_1,x_2}+f'_{x_1,x_2}
\right)\psi=k^2 \psi.
\end{equation}
As shown in appendix~A, this Schr\"odinger equation has two independent solutions that behave outside the interval $(x_1,x_2)$ as\footnote{
%
%
In some of the previous papers, there is an error in the equations corresponding to (\ref{2-2.10}) (equations (2.3) of \cite{low}, equations (1.6) of \cite{analysis}, and equations (3.2) of \cite{expressions}).
The functions defined in these equations are solutions of the Schr\"odinger equation, and not the Fokker-Planck equation. For these functions to become solutions of the Fokker-Planck equation, there should be a factor $\rme^{[V(x_2)-V(x_1)]/2}$ or $\rme^{-[V(x_2)-V(x_1)]/2}$ in front of the transmission coefficient $\tau(x_2,x_1;k)$.
 This error does not affect any of the results of these papers.
}
\refstepcounter{equation}
\label{2-2.10}
\addtocounter{equation}{-1}
\numparts
\begin{equation}
\label{2-2.10a}
\psi_1(x)=
\cases{
\tau(x_2,x_1;k)\, \rme^{-\rmi k(x-x_1)} & $(x<x_1)$ \\
\rme^{-\rmi k(x-x_2)} + R_r(x_2,x_1;k)\, \rme^{\rmi k(x-x_2)} & $(x>x_2)$, \\
}
\end{equation}
\begin{equation}
\label{2-2.10b}
\psi_2(x)=
\cases{
e^{\rmi k(x-x_1)} + R_l(x_2,x_1;k)\, \rme^{-\rmi k(x-x_1)} &$(x<x_1)$ \\
\tau(x_2,x_1;k)\, \rme^{\rmi k(x-x_2)} &  $(x>x_2)$. \\
}
\end{equation}
\endnumparts
Thus, $R_r(x_2,x_1;k)$ (or $R_l(x_2,x_1;k)$) is the factor multiplying the wave reflected from the interval $(x_1,x_2)$ when an incident wave is coming into this interval from the right (or, respectively, left), and $\tau(x_2,x_1;k)$ is the factor multiplying the transmitted wave.

We define
\begin{equation}
\label{1-2.8}
S_r(x,k) \equiv \frac{R_r(x,-\infty;k)}{1+R_r(x,-\infty;k)}, 
\qquad
S_l(x,k) \equiv \frac{R_l(\infty;x;k)}{1+R_l(\infty,x;k)},
\end{equation}
\begin{equation}
\label{1-2.9}
S(x,k)\equiv S_r(x,k)+S_l(x,k).
\end{equation}
The reflection coefficients for semi-infinite intervals, which appear in (\ref{1-2.8}),  are well defined as long as ${\rm Im}\,k>0$. (The well-definedness of $R_r(x,-\infty;k)$ for periodic potentials is explained in appendix~B. For non-periodic potentials, see appendix~G of \cite{low}. A related discussion is found in \cite{r5}.) For ${\rm Im}\,k=0$, we define them as $R_r(x,-\infty;k)\equiv \lim_{\epsilon \downarrow 0}R_r(x,-\infty;k+\rmi \epsilon)$ and $R_l(\infty,x;k)\equiv \lim_{\epsilon \downarrow 0}R_l(\infty,x;k+\rmi \epsilon)$. 
As explained in appendix~C, the functions $S_r$, $S_l$ and $S$ are closely related to the Weyl-Titchmarsh $m$-function.
The Green function can be expressed in terms of this $S$ as \cite{expressions}
\begin{equation}
\label{1-2.10}
\fl
G_{\rm S}(x,y;k)=
\frac{1}{2\rmi k\sqrt{[1-S(x,k)][1-S(y,k)]}}
\exp\left[
\rmi k(x-y)-\rmi k\int_y^x S(z,k)\,\rmd z
\right].
\end{equation}

In our formalism, we deal with the scattering coefficients $\tau$, $R_r$ and $R_l$ in an generalized form with an additional variable $W$. First, we define $\bar \alpha$ and $\bar \beta$ by
\begin{eqnarray}
\label{1-2.11}
\fl
\left(
\begin{array}{cc}
\bar \alpha(x,x';W;k) & \bar \beta(x,x';W;-k) \\
\bar \beta(x,x';W;k) & \bar \alpha(x,x';W;-k) \\
\end{array}
\right) 
\nonumber \\
\fl \qquad \qquad
\equiv
\left(
\begin{array}{cc}
\cosh\frac{W-V(x)}{2} &  -\sinh\frac{W-V(x)}{2} \\
-\sinh\frac{W-V(x)}{2} & \cosh\frac{W-V(x)}{2} \\
\end{array}
\right)
\left(
\begin{array}{cc}
\alpha(x,x';k) & \beta(x,x';-k) \\
\beta(x,x';k) & \alpha(x,x';-k) \\
\end{array}
\right).
\end{eqnarray}
(Here, as elsewhere in this paper, the bar does not denote complex conjugation.)
Then, $\bar \tau$, $\bar R_r$ and $\bar R_l$ are defined in the same way as (\ref{1-2.7}) with $\alpha\to \bar \alpha$ and $\beta \to \bar \beta$:
\refstepcounter{equation}
\label{1-2.12}
\addtocounter{equation}{-1}
\numparts
\begin{equation}
\label{1-2.12a}
\bar \tau(x,x';W;k)\equiv 
\frac{1}{\bar \alpha(x,x';W;k)}
=\frac{\sqrt{1-\xi^2(x,W)}\, \tau(x,x';k)}{1-\xi(x,W)R_r(x,x';k)},
\end{equation}
\begin{equation}
\label{1-2.12b}
\bar R_r(x,x';W;k)\equiv 
\frac{\bar \beta(x,x';W;k)}{\bar \alpha(x,x';W;k)}
=\frac{R_r(x,x';k)-\xi(x,W)}{1-\xi(x,W)R_r(x,x';k)},
\end{equation}
\begin{equation}
\fl
\bar R_l(x,x';W;k)\equiv 
-\frac{\bar \beta(x,x';W;-k)}{\bar \alpha(x,x';W;k)}
=R_l(x,x';k) + \frac{\xi(x,W)\tau^2(x,x';k)}{1-\xi(x,W)R_r(x,x';k)},
\end{equation}
\endnumparts
where
\begin{equation}
\label{1-2.13}
\xi(x,W)\equiv \tanh\frac{W-V(x)}{2}.
\end{equation}
The meaning of these generalized scattering coefficients is explained in \cite{algebraic}.
The original scattering coefficients $\tau$, $R_r$ and $R_l$ are recovered from $\bar \tau$, $\bar R_r$ and $\bar R_l$ by setting $W=V(x)$.

For $k=0$, equation (\ref{1-2.1}) can be exactly solved. 
It is easy to see that 
\begin{equation}
\fl
\alpha(x,x';0)=\cosh\frac{V(x')-V(x)}{2}, \qquad 
\beta(x,x';0)=\sinh\frac{V(x')-V(x)}{2}.
\end{equation}
Hence we have the expressions of $\bar \tau$, $\bar R_r$ and $\bar R_l$ for $k=0$ as
\begin{eqnarray}
\label{1-2.14}
\bar \tau(x,x';W;0)={\rm sech}\, \frac{W-V(x')}{2},
\nonumber \\
\fl
\bar R_r(x,x';W;0)=-\tanh\frac{W-V(x')}{2}, \qquad 
\bar R_l(x,x';W;0)=\tanh\frac{W-V(x')}{2}.
\end{eqnarray}

For $n=1,2,3,\ldots$ and $-\infty\leq a\leq b\leq \infty$,
we define the notation
\begin{equation}
\label{1-2.15}
\fl
[\sigma_1,\sigma_2,\ldots,\sigma_n]_a^b\equiv \int \cdots \int_{a \leq z_1\leq z_2 \leq \cdots \leq z_n\leq b}
\rmd z_1 \rmd z_2\cdots \rmd z_n
\exp\Biggl[
\sum_{j=1}^n \sigma_j V(z_j)
\Biggr],
\end{equation}
where each $\sigma_j$ is either $+1$ or $-1$.
For simplicity, we use the obvious abbreviation $[\hbox{$+$}]_a^b$, $[\hbox{$-$}]_a^b$,  $[\hbox{$+$}\hbox{$-$}]_a^b$, etc in place of $[+1]_a^b$, $[-1]_a^b$, $[+1,-1]_a^b$, etc.  
The integrals of the form of (\ref{1-2.15}) satisfy the multiplication rule
\begin{eqnarray}
\label{1-2.16}
\fl
[\sigma_1,\sigma_2,\ldots,\sigma_n]_a^b\times [\sigma']_a^b
&=[\sigma',\sigma_1,\sigma_2,\ldots,\sigma_n]_a^b
+[\sigma_1, \sigma', \sigma_2,\ldots,\sigma_n]_a^b
+[\sigma_1,\sigma_2, \sigma' \ldots,\sigma_n]_a^b
\nonumber \\
\fl
&\qquad +\cdots + [\sigma_1,\sigma_2,\ldots,\sigma', \sigma_n]_a^b
+[\sigma_1,\sigma_2,\ldots,\sigma_n,\sigma']_a^b.
\end{eqnarray}
When the potential satisfies (\ref{1-1.8}), the quantities defined by
\begin{equation}
\label{1-2.17}
M\equiv [\hbox{$-$}]_{x-L}^x=\int_{x-L}^x\rme^{-V(z)}\rmd z, \qquad 
P \equiv [\hbox{$+$}]_{x-L}^x=\int_{x-L}^x\rme^{V(z)}\rmd z
\end{equation}
are independent of $x$.
It is also convenient to define
\begin{equation}
\label{1-2.18}
L_0 \equiv \sqrt{P M}, \qquad \rme^{V_0}\equiv \sqrt{P/M}, 
\quad {\rm i.e.}\quad V_0 \equiv (1/2)\log(P/M).
\end{equation}
From (\ref{1-2.16}) and (\ref{1-2.17}), we have
\begin{equation}
\label{1-2.19}
\fl
[\hbox{$+$$+$}]_{x-L}^x=P^2/2, \qquad
[\hbox{$-$$-$}]_{x-L}^x=M^2/2, \qquad
[\hbox{$+$$-$}]_{x-L}^x+\,[\hbox{$-$$+$}]_{x-L}^x=P M=L_0^2.
\end{equation}
We can check that the left-hand sides of (\ref{1-2.19}) are independent of $x$ by differentiating them with respect to $x$, using
\begin{equation}
\label{1-2.20}
\fl
\frac{\rmd}{\rmd x}[\sigma_1,\sigma_2,\ldots,\sigma_n]_{x-L}^x
=[\sigma_1,\sigma_2,\ldots,\sigma_{n-1}]_{x-L}^x \,\rme^{\sigma_n V(x)}
-[\sigma_2,\sigma_3,\ldots,\sigma_n]_{x-L}^x \,\rme^{\sigma_1V(x)}.
\end{equation}

We define the operators $\mathcal{A}$ and $\mathcal{B}$, which act on functions of $x$ and $W$, as
\begin{eqnarray}
\label{1-2.21}
\mathcal{A}h(x,W) &\equiv \frac{\partial}{\partial x}h(x,W),
\\
\label{1-2.22}
\mathcal{B}h(x,W) 
&\equiv \frac{\partial}{\partial W}
\left\{ \sinh[W-V(x)]h(x,W)\right\}
\nonumber \\
&=
\left\{ \cosh[W-V(x)]+\sinh[W-V(x)]\frac{\partial}{\partial W}
\right\}h(x,W).
\end{eqnarray}
We assume that the functions $h(x,W)$, on which these operators act, are (i) piecewise continuously differentiable\footnote{
%
%
Here we do not require $h(x,W)$ to be continuous in $x$. We allow $h$ to have jumps, and so $\partial h/\partial x$ may contain delta functions.
The argument in \cite{analysis} is valid without the requirement of continuity.
}
 with respect to $x$, and (ii) analytic with respect to $W$ on the real axis.

Finally, the operator $\mathcal{D}$ is defined as
\begin{equation}
\label{1-2.23}
\mathcal{D}h(x,W) \equiv h(x,W)-h(x,V_0),
\end{equation}
with $V_0$ defined by (\ref{1-2.18}). In terms of differential and integral operators, we can write
\begin{equation}
\label{1-2.24}
\mathcal{D}=\int_{V_0}^W \rmd W \frac{\partial}{\partial W}.
\end{equation}


\section{The low-energy expansion formula for non-periodic potentials}
Here we review the expansion formula for reflection coefficients derived in \cite{analysis}, where the potential $V(x)$ was assumed to be monotone for sufficiently large $\vert x \vert$. For details, see~\cite{analysis} and \cite{low}. In this section, $V(x)$ is not assumed to be periodic.

With expression (\ref{1-2.10}), the analysis of the Green function is reduced to the analysis of the reflection coefficients $R_r(x,-\infty;k)$ and $R_l(\infty,x;k)$. 
Here we consider $R_r(x,-\infty;k)$. 
As a function of $x$, the reflection coefficient satisfies a nonlinear differential equation of Riccati type. By introducing the generalized reflection coefficient (\ref{1-2.12b}) with an additional variable $W$, we can turn this nonlinear equation into a linear partial differential equation for the two variables $x$ and $W$. This equation has the form
\begin{equation}
\label{1-3.1}
(\mathcal{A}-2\rmi k\mathcal{B})
\left[\bar R_r(x,-\infty;W;k)+\xi(x,W)\right]
=[1-\xi^2(x,W)]f(x)
\end{equation}
with the operators $\mathcal{A}$ and $\mathcal{B}$ defined by (\ref{1-2.21}) and (\ref{1-2.22}), respectively.
The reflection coefficient can be obtained by solving (\ref{1-3.1}). 
Solving the differential equation (\ref{1-3.1}) with an appropriate boundary condition is equivalent to finding the inverse of the operator $\mathcal{A}-\rmi k \mathcal{B}$ with an appropriately restricted domain. 

Let us start with the case $k=0$. 
If we restrict the domain of the operator $\mathcal{A}$ to functions satisfying
\begin{equation}
\label{1-3.2}
\lim_{x \to -\infty}h(x,W)=0,
\end{equation}
then we have the inverse of $\mathcal{A}$ as
\begin{equation}
\label{1-3.3}
\mathcal{A}^{-1}g(x,W)=\int_{-\infty}^x g(z,W)\,\rmd z,
\end{equation} 
where $g(x,W)=(\partial/\partial x)h(x,W)$ with some $h$ satisfying (\ref{1-3.2}).
The operator $\mathcal{A}^{-1}$ given by (\ref{1-3.3}) satisfies $\mathcal{A}^{-1}\mathcal{A}h=h$ and $\mathcal{A}\mathcal{A}^{-1}g=g$ if $h$ satisfies (\ref{1-3.2}).

This can be extended to $k\neq 0$.
We restrict the domain of the operator $\mathcal{A}-\rmi k \mathcal{B}$ to functions $h(x,W)$ which satisfy
\begin{equation}
\label{1-3.4}
\lim_{z \to -\infty} \frac{\bar \tau^2(x,z;W;k)}{1-\bar R_l^2(x,z;W;k)} h(z,\bar \omega(x,z;W;k))=0
\end{equation} 
for any $x$, where
\begin{equation}
\label{1-3.5}
\bar \omega(x,z;W;k) \equiv V(z)+\log \frac{1+\bar R_l(x,z;W;k)}{1-\bar R_l(x,z;W;k)}.
\end{equation}
Then the inverse of $\mathcal{A}-\rmi k \mathcal{B}$ is given by
\begin{equation}
\label{1-3.6}
\fl
(\mathcal{A}-2 \rmi k\mathcal{B})^{-1} g(x,W)
=\int_{-\infty}^x \rmd z \frac{\bar \tau^2(x,z;W;k)}{1-\bar R_l^2(x,z;W;k)}
g(z,\bar \omega(x,z;W;k)).
\end{equation}
This $(\mathcal{A}-\rmi k \mathcal{B})^{-1}$ satisfies $(\mathcal{A}-\rmi k \mathcal{B})^{-1}(\mathcal{A}-\rmi k \mathcal{B})h=h$ and $(\mathcal{A}-\rmi k \mathcal{B})(\mathcal{A}-\rmi k \mathcal{B})^{-1}g=g$ provided that $h$ satisfies (\ref{1-3.4}).
In (\ref{1-3.4}), the expression $h(z,\bar \omega(x,z;W;k))$ stands for the quantity obtained from $h(x,W)$ by inserting $z$ and $\bar \omega(x,z;W;k)$ in place of $x$ and $W$, respectively. 
From (\ref{1-2.14}) and (\ref{1-3.5}), we have $\bar \omega(x,z;W;0)=W$. By using this and (\ref{1-2.14}), we can see that (\ref{1-3.4}) reduces to (\ref{1-3.2}) if we set $k=0$. Similarly, the right-hand side of (\ref{1-3.6}) reduces to (\ref{1-3.3}) when $k$ is set to be zero.
 
It can be shown that $\bar R_r(x,-\infty;W;k)+\xi(x,W)$, as a function of $x$ and $W$, satisfies (\ref{1-3.4}). So it is possible to apply $(\mathcal{A}-2\rmi k\mathcal{B})^{-1}$ to both sides of (\ref{1-3.1}) and obtain
\begin{equation}
\label{1-3.7}
\bar R_r(x,-\infty;W;k)
=-\xi+(\mathcal{A}-2\rmi k\mathcal{B})^{-1}(1-\xi^2)f.
\end{equation}
By iterating the identity
\begin{equation}
\label{1-3.8}
(\mathcal{A}-2 \rmi k\mathcal{B})^{-1}=\mathcal{A}^{-1} + 2 \rmi k (\mathcal{A}-2 \rmi k\mathcal{B})^{-1} \mathcal{B}\mathcal{A}^{-1},
\end{equation}
we have the expression
\begin{eqnarray}
\label{1-3.9}
(\mathcal{A}-2 \rmi k\mathcal{B})^{-1}&=
\left[1+ \rmi k \mathcal{L} + (\rmi k)^2 \mathcal{L}^2 + \cdots + (\rmi k)^N\mathcal{L}^N\right]\mathcal{A}^{-1}
\nonumber \\
&\qquad +(\rmi k)^{N+1}(\mathcal{A}-2 \rmi k\mathcal{B})^{-1}\mathcal{A}\mathcal{L}^{N+1} \mathcal{A}^{-1},
\end{eqnarray}
where
\begin{equation}
\label{1-3.10}
\mathcal{L} \equiv 2\mathcal{A}^{-1} \mathcal{B}.
\end{equation}
The expansion of $\bar R_r(x,-\infty;W;k)$ in powers of $k$ can be obtained by substituting (\ref{1-3.9}) into (\ref{1-3.7}). We have
\begin{equation}
\label{1-3.11}
\bar R_r=\bar r_0 + \rmi k \bar r_1 + (\rmi k)^2 \bar r_2 + \cdots + (\rmi k)^N \bar r_N+\bar \rho_N
\end{equation}
with the coefficients of the expansion
\refstepcounter{equation}
\label{1-3.12}
\addtocounter{equation}{-1}
\numparts
\begin{equation}
\label{1-3.12a}
\bar r_0=\mathcal{A}^{-1}(1-\xi^2)f - \xi,
\end{equation}
\begin{equation}
\label{1-3.12b}
\bar r_n=\mathcal{L}^n(\bar r_0 +\xi) \qquad (n \geq 1),
\end{equation}
\endnumparts
and the remainder term
\begin{equation}
\label{1-3.13}
\bar \rho_N =(\rmi k)^{N+1}(\mathcal{A}-2 \rmi k\mathcal{B})^{-1}\mathcal{A}\,\bar r_{N+1}.
\end{equation}

This method works well if $V(x)$ either converges to a definite value or diverges to $\pm \infty$ as $x \to -\infty$. 
From (\ref{1-1.2}) and (\ref{1-2.13}) we find
\begin{equation}
\label{1-3.14}
(1-\xi^2) f
=\frac{-1}{2 \cosh^2 \frac{W-V(x)}{2}}\frac{\rmd}{\rmd x} V(x)
=\frac{\partial}{\partial x} \tanh\frac{W-V(x)}{2}.
\end{equation}
Substituting this into (\ref{1-3.3}) gives
\begin{equation}
\label{1-3.15}
\fl
\mathcal{A}^{-1} (1-\xi^2) f=\int_{-\infty}^x
\left(\frac{\partial}{\partial z} \tanh\frac{W-V(z)}{2}\right)\rmd z
=\xi-\tanh\frac{W-V(-\infty)}{2}.
\end{equation}
The zeroth-order coefficient of the expansion (equation~(\ref{1-3.12a})) is thus
\begin{equation}
\label{1-3.16}
\bar r_0=-\tanh \frac{W-V(-\infty)}{2}.
\end{equation}
If $V(-\infty)=\pm \infty$, then $\bar r_0=\pm 1$. 
The higher order coefficients can be obtained by substituting (\ref{1-3.16}) into (\ref{1-3.12b}). 
The resulting expressions are given in \cite{analysis} and \cite{low}.


\section{Application of the expansion formula to periodic potentials}
The expansion formula shown in the previous section cannot be directly applied to periodic potentials. 
If $V$ is periodic, the right-hand side of (\ref{1-3.16}) does not have a definite value. 
It is the objective of this paper to make a modification to this formula so that it becomes applicable to the periodic case. 

Equation (\ref{1-3.1}) holds for any potential, whether it is periodic or not. 
The function $h(x,W)=\bar R_r(x,-\infty;W;k)+\xi(x,W)$ is periodic in $x$ when the potential is periodic. If $h$ is periodic, then (\ref{1-3.4}) is satisfied as long as ${\rm Im}\,k>0$.
Therefore, the basic expressions (\ref{1-3.7}) and (\ref{1-3.6}) are correct even when the potential is periodic.  
 (When ${\rm Im}\,k=0$, it is necessary to replace $k$ in the right-hand side of (\ref{1-3.6}) by $k+\rmi \epsilon$ with $\epsilon>0$, and take the limit $\epsilon \to 0$ after evaluating the integral.)
 
If we try to apply the method of section~3 to a periodic potential, we need to let $\mathcal{A}^{-1}$ act on periodic functions, as in equations (\ref{1-3.12a}) and (\ref{1-3.12b}). 
However, since (\ref{1-3.2}) is not satisfied when $h(x,W)$ is periodic, we cannot use  (\ref{1-3.3}) for periodic functions. The right-hand side of (\ref{1-3.3}) is indeterminate when $g(x,W)$ is periodic in $x$. 

Thus, in order to deal with periodic potentials, we must discard (\ref{1-3.3}) and find another appropriate expression for the inverse of $\mathcal{A}$. 
Since (\ref{1-3.6}) is valid for both periodic and non-periodic cases, we can derive $\mathcal{A}^{-1}$ from (\ref{1-3.6}) as
\begin{eqnarray}
\label{1-4.1}
\mathcal{A}^{-1} g(x,W) &= \lim_{k \to 0}(\mathcal{A}-2 \rmi k\mathcal{B})^{-1} g(x,W)
\nonumber \\
&=\lim_{k \to 0}\int_{-\infty}^x  \frac{\bar \tau^2(x,z;W;k)}{1-\bar R_l^2(x,z;W;k)}
g(z,\bar \omega(x,z;W;k))\,\rmd z.
\end{eqnarray}
As we shall see later, this gives the correct inverse of $\mathcal{A}$ even when the potential is periodic.
This expression includes (\ref{1-3.3}) as a special case.
As noted below equation (\ref{1-3.6}), the right-hand side of (\ref{1-3.3}) is obtained by letting $k \to 0$ inside the integral, i.e.
\begin{equation}
\label{1-4.2}
\fl
\int_{-\infty}^x  \lim_{k \to 0}\frac{\bar \tau^2(x,z;W;k)}{1-\bar R_l^2(x,z;W;k)}g(z,\bar \omega(x,z;W;k))\,\rmd z=\int_{-\infty}^x g(z,W)\,\rmd z.
\end{equation}  
Hence, we find that (\ref{1-4.1}) reduces to (\ref{1-3.3}) if the limit and the integral in (\ref{1-4.1}) are interchangeable. This is what happens in the non-periodic cases discussed in the previous papers.
In the periodic case, however, the limit and the integral cannot be interchanged, and so  the correct inverse of $\mathcal{A}$ is different from (\ref{1-3.3}).

What we mean by ^^ ^^ correct inverse of $\mathcal{A}$" is the operator $\mathcal{A}^{-1}$ that can be used in equations (\ref{1-3.12}) to produce the correct expansion of $\bar R_r$. 
As we have not yet specified the domain of $\mathcal{A}$, the term ^^ ^^ inverse of $\mathcal{A}$" does not yet have a definite meaning. For the time being, let us accept (\ref{1-4.1}) as the definition of $\mathcal{A}^{-1}$. It will be shown in section~6, after properly defining the domain of $\mathcal{A}$, that this $\mathcal{A}^{-1}$ is indeed the inverse of $\mathcal{A}$.

The method of section~3 remains valid for periodic potentials if (\ref{1-3.3})  is is replaced by (\ref{1-4.1}).
Our main task is to calculate the right-hand side of (\ref{1-4.1}) to derive an expression of $\mathcal{A}^{-1}$ for the periodic case. This will be done in the next section. The result is
\begin{equation}
\label{1-4.3}
\fl
\mathcal{A}^{-1}g(x,W)=
\frac{1}{L_0 \sinh(W-V_0)}
\,\mathcal{D}\int \!\!\!\int_{x-L\leq z \leq z'\leq x} \rmd z\, \rmd z'\sinh[V(z')-W] g(z,W),
\end{equation}
where $L_0$, $V_0$ and $\mathcal{D}$ are defined by (\ref{1-2.18}) and (\ref{1-2.23}).
More explicitly, this means
\begin{eqnarray}
\label{1-4.4}
\fl
\mathcal{A}^{-1}g(x,W)=
\frac{1}{L_0 \sinh(W-V_0)}
\int_{x-L}^x \rmd z \int_z^x\rmd z'
\Bigl\{& \sinh[V(z')-W] g(z,W)
\nonumber \\
& \quad -\sinh[V(z')-V_0]g(z,V_0) \Bigr\}.
\end{eqnarray}
This, instead of (\ref{1-3.3}),  is the inverse of $\mathcal{A}$ appropriate for periodic potentials.

Hereafter we shall always assume that $V(x)$ is periodic. The symbol $\mathcal{A}^{-1}$ shall always refer to (\ref{1-4.1}) or (\ref{1-4.3}), and not (\ref{1-3.3}).


\section{Derivation of (\ref{1-4.3})}
In this section, we shall derive (\ref{1-4.3}) from (\ref{1-4.1}).
We assume that $g(x,W)=\mathcal{A}h(x,W)$ with some function $h$ satisfying $h(x+L,W)=h(x,W)$. Then it follows that
\begin{equation}
\label{1-5.1}
g(x+L,W)=g(x,W),
\end{equation}
\begin{equation}
\label{1-5.2}
\int_{x-L}^x g(z,W)\,\rmd z=0,
\end{equation}
where $x$ is arbitrary.  As it turns out, (\ref{1-4.1}) does not make sense unless (\ref{1-5.2}) is satisfied.

The expressions appearing in (\ref{1-3.6}) and (\ref{1-3.5}) can be written as
\refstepcounter{equation}
\label{2-5.20}
\addtocounter{equation}{-1}
\numparts
\begin{eqnarray}
\frac{\bar \tau^2(x,z;W;k)}{1-\bar R_l^2(x,z;W;k)}
=\frac{1}{\bar \alpha^2(x,z;W;k)-\bar \beta^2(x,z;W;-k)},
\\
\frac{1+\bar R_l(x,z;W;k)}{1-\bar R_l(x,z;W;k)}
=\frac{\bar \alpha(x,z;W;k)-\bar \beta(x,z;W;-k)}{\bar \alpha(x,z;W;k)+\bar \beta(x,z;W;-k)}.
\end{eqnarray}
\endnumparts
So, in order to calculate the right-hand side of (\ref{1-4.1}), we need to know the small-$k$ behavior of $\bar \alpha(x,z;W;k)$ and $\bar \beta(x,z;W;-k)$. By definition, $\bar \alpha(x,z;W;k)$ and $\bar \beta(x,z;W;-k)$ are obtained from the elements of the matrix $U(x,z;k)$. 

We divide the domain of the integral in (\ref{1-4.1}) into regions of length $L$ as
\begin{equation}
\label{1-5.3}
\int_{-\infty}^x = \sum_{n=0}^\infty \int_{x-(n+1)L}^{x-nL}.
\end{equation}
To deal with each term on the right-hand side of (\ref{1-5.3}), we assume that
\begin{equation}
\label{1-5.4}
 x-(n+1)L < z \leq x-nL.
\end{equation}
Using (\ref{1-2.5}) and (\ref{1-2.6}), the matrix $U(x,z;k)$ is factorized as
\begin{equation}
\label{1-5.5}
\fl
U(x,z;k)=U(x,x-nL;k) U(x-nL,z;k)=[U(x,x-L;k)]^n U(x-nL,z;k).
\end{equation}
We can calculate $[U(x,x-L;k)]^n$, which appears in (\ref{1-5.5}), by diagonalizing $U(x,x-L;k)$.
The eigenvalues of $U(x,x-L;k)$ are $\lambda(k)$ and $1/\lambda(k)$, where
\begin{equation}
\label{1-5.6}
\lambda(k) \equiv Y(k)-\rmi \sqrt{1-[Y(k)]^2},
\end{equation}
\begin{equation}
\label{1-5.7}
Y(k) \equiv \frac{1}{2}[\alpha(x,x-L;k)+\alpha(x,x-L;-k)].
\end{equation}
We used (\ref{1-2.4}) to derive (\ref{1-5.6}). 
As shown in appendix~D, the quantity $Y(k)$ is independent of $x$, and so is $\lambda(k)$. 
The branch of the square root in (\ref{1-5.6}) is chosen so that $\vert \lambda(k) \vert>1$ for ${\rm Im}\,k>0$.
The two eigenvalues of $U$ are the reciprocal of each other, reflecting the fact that the determinant of $U$ is unity. Also note that $\lambda(-k)=1/\lambda(k)$.
It is straightforward to show that $[U(x,x-L;k)]^n$ is expressed in terms of $\lambda$ as
\begin{eqnarray}
\label{1-5.8}
\fl
[U(x,x-L)]^n=
\frac{1}{\lambda-\lambda^{-1}}
\left(
\begin{array}{cc}
\lambda^{-n}(\lambda-\alpha) -\lambda^n(\lambda^{-1}-\alpha) & 
\beta' (\lambda^n-\lambda^{-n}) \nonumber \\
\beta (\lambda^n-\lambda^{-n}) &
\lambda^n (\lambda-\alpha) -\lambda^{-n}(\lambda^{-1}-\alpha) \cr
\end{array}
\right),
\nonumber \\
\fl
\end{eqnarray}
where we have omitted the dependence on $k$, and used the abbreviation
\begin{equation}
\label{1-5.9}
\fl
\alpha\equiv \alpha(x,x-L;k), \qquad \beta \equiv \beta(x,x-L;k), \qquad 
\beta' \equiv \beta(x,x-L;-k).
\end{equation}

In general, for any finite $x_1$ and $x_2$, we can expand $\alpha(x_2,x_1;k)$ and $\beta(x_2,x_1;k)$  in powers of $k$ as
\refstepcounter{equation}
\label{1-5.10}
\addtocounter{equation}{-1}
\numparts
\begin{eqnarray}
\label{1-5.10a}
\fl
\alpha(x_2,x_1;k)=\cosh \case{V(x_1)-V(x_2)}{2}
-\frac{\rmi k}{2}
\left(\rme^{[V(x_1)+V(x_2)]/2}\,[\hbox{$-$}]_{x_1}^{x_2}+\rme^{-[V(x_1)+V(x_2)]/2}\,[\hbox{$+$}]_{x_1}^{x_2}\right)
\nonumber \\
 -\frac{k^2}{2}
\left(\rme^{[V(x_1)-V(x_2)]/2}\,[\hbox{$-$}\hbox{$+$}]_{x_1}^{x_2}+\rme^{[V(x_2)-V(x_1)]/2}\,[\hbox{$+$}\hbox{$-$}]_{x_1}^{x_2}\right)+O(k^3),
\end{eqnarray}
\begin{eqnarray}
\label{1-5.10b}
\fl
\beta(x_2,x_1;k)=\sinh\case{V(x_1)-V(x_2)}{2} 
+\frac{\rmi k}{2}
\left(\rme^{[V(x_1)+V(x_2)]/2}\,[\hbox{$-$}]_{x_1}^{x_2}-\rme^{-[V(x_1)+V(x_2)]/2}\,[\hbox{$+$}]_{x_1}^{x_2}\right)
\nonumber \\
 -\frac{k^2}{2}
\left(\rme^{[V(x_1)-V(x_2)]/2}\,[\hbox{$-$}\hbox{$+$}]_{x_1}^{x_2}-\rme^{[V(x_2)-V(x_1)]/2}\,[\hbox{$+$}\hbox{$-$}]_{x_1}^{x_2}\right)+O(k^3).
\end{eqnarray}
\endnumparts
The derivation of (\ref{1-5.10}) is given in appendix~B. 
With $x_2=x$ and $x_1=x-L$, equation (\ref{1-5.10a}) reads
\begin{eqnarray}
\label{1-5.11}
\alpha(x,x-L;k)&=1-\frac{\rmi k}{2}\left(\rme^{V(x)}M +\rme^{-V(x)}P \right) -\frac{k^2}{2}P M+O(k^3)
\nonumber \\
&=1-\rmi k L_0 \cosh[V(x)-V_0] -\frac{k^2}{2}L_0^2+O(k^3),
\end{eqnarray}
where we have used (\ref{1-2.18}) and (\ref{1-2.19}).
Substituting (\ref{1-5.11}) into (\ref{1-5.7}) and (\ref{1-5.6}), we find
\begin{equation}
\label{1-5.12}
Y(k)=1-\frac{k^2}{2}L_0^2+O(k^4), \qquad
\sqrt{1-[Y(k)]^2}=k L_0+O(k^3),
\end{equation}
\begin{equation}
\label{1-5.13}
\fl
\lambda(k)=1-\rmi k L_0 -\frac{k^2}{2}L_0^2+O(k^3), \qquad [\lambda(k)]^{-1}=1+\rmi k L_0 -\frac{k^2}{2}L_0^2+O(k^3).
\end{equation}
Hence, the small-$k$ expressions of the quantities appearing in (\ref{1-5.8}) are
\refstepcounter{equation}
\label{1-5.14}
\addtocounter{equation}{-1}
\numparts
\begin{equation}
\lambda-\lambda^{-1}=-2\rmi k L_0+O(k^3),
\end{equation}
\begin{equation}
\lambda-\alpha(x,x-L;k)=2 \rmi k L_0\sinh^2\frac{V(x)-V_0}{2}+ O(k^3),
\end{equation}
\begin{equation}
\lambda^{-1}-\alpha(x,x-L;k)=2 \rmi k L_0\cosh^2\frac{V(x)-V_0}{2}+ O(k^3).
\end{equation}
\endnumparts
The expressions for $\beta$ and $\beta'$ are given by (\ref{1-5.10b}) as
\begin{equation}
\label{1-5.15}
\fl
\beta(x,x-L;\pm k)= \pm \rmi k L_0 \sinh[V(x)-V_0]
-\frac{k^2}{2}\left(\,[\hbox{$-$}\hbox{$+$}]_{x-L}^{x}-\,[\hbox{$+$}\hbox{$-$}]_{x-L}^{x}\right)+O(k^3).
\end{equation}
By substituting (\ref{1-5.14}) and (\ref{1-5.15}) into (\ref{1-5.8}), we obtain the elements of $U(x,x-n L;k)=[U(x,x-L;k)]^n$ as
\refstepcounter{equation}
\label{2-5.16}
\addtocounter{equation}{-1}
\numparts
\begin{eqnarray}
\fl
\alpha(x,x-nL;\pm k)&=\cosh^2 \case{V(x)-V_0}{2}\, \lambda^{\pm n}
-\sinh^2 \case{V(x)-V_0}{2}\, \lambda^{\mp n} +O(k^2),
\\
\fl
\beta(x,x-nL;\pm k)&=
\mp \cosh \case{V(x)-V_0}{2} \sinh \case{V(x)-V_0}{2} \left(\lambda^n-\lambda^{-n}\right)
\nonumber \\
& \qquad - \frac{\rmi k}{4 L_0}\left(\,[\hbox{$-$}\hbox{$+$}]_{x-L}^{x}-\,[\hbox{$+$}\hbox{$-$}]_{x-L}^{x}\right)\left(\lambda^n-\lambda^{-n}\right)+O(k^2).
\end{eqnarray}
\endnumparts
(In these equations, $\lambda$ stands for $\lambda(k)$ and not $\lambda(\pm k)$. Recall that $\lambda(-k)=1/\lambda(k)$.)
Since we are going to take the sum over $n$ as in (\ref{1-5.3}), we cannot expand $\lambda^n$ and $\lambda^{-n}$ in powers of $k$ and neglect the higher order terms. Even if $\vert k \vert$ is small, the higher order terms are not negligible when $n$ is large.
So, we leave $\lambda^n$ and $\lambda^{-n}$ as they are.

Up to order $k$, the elements of $U(x-nL,z;k)$ are given by (\ref{1-5.10}) as
\refstepcounter{equation}
\label{2-5.17}
\addtocounter{equation}{-1}
\numparts
\begin{eqnarray}
\fl
\alpha(x-nL,z;\pm k)&=\cosh \case{V(z)-V(x)}{2}
\mp\frac{\rmi k}{2}
\left(\rme^{[V(x)+V(z)]/2}\,[\hbox{$-$}]_{z}^{x-nL}+\rme^{-[V(x)+V(z)]/2}\,[\hbox{$+$}]_{z}^{x-nL}\right)
\nonumber \\
& \qquad
+O(k^2),
\end{eqnarray}
\begin{eqnarray}
\fl
\beta(x-nL,z;\pm k)&=\sinh\case{V(z)-V(x)}{2} 
\pm \frac{\rmi k}{2}
\left(\rme^{[V(x)+V(z)]/2}\,[\hbox{$-$}]_{z}^{x-nL}-\rme^{-[V(x)+V(z)]/2}\,[\hbox{$+$}]_{z}^{x-nL}\right)
\nonumber \\
& \qquad
+O(k^2).
\end{eqnarray}
\endnumparts
The expressions of $\alpha(x,z;\pm k)$ and $\beta(x,z;\pm k)$ are obtained by substituting (\ref{2-5.16}) and (\ref{2-5.17}) into $U(x,z;k)=U(x,x-nL;k) U(x-nL,z;k)$, or, explicitly,
\begin{eqnarray}
\label{2-5.18}
\fl
\alpha(x,z;\pm k)
&=\alpha(x,x-nL;\pm k)\alpha(x-nL,z;\pm k)
+\beta(x,x-nL;\mp k)\beta(x-nL,z;\pm  k),
\nonumber \\
\fl
\beta(x,z;\pm k)
&=\beta(x,x-nL;\pm k)\alpha(x-nL,z;\pm k)
+\alpha(x,x-nL;\mp k)\beta(x-nL,z;\pm  k).
\nonumber \\
\fl
\end{eqnarray}
Substituting the resulting expressions into the definition of $\bar \alpha$ and $\bar \beta$ (equations (\ref{1-2.11})) yields
\refstepcounter{equation}
\label{1-5.17}
\addtocounter{equation}{-1}
\numparts
\begin{eqnarray}
\fl
\bar \alpha(x,z;W;k)=
\left(\cosh \case{V(z)-V_0}{2}\cosh \case{W-V_0}{2}\right)\lambda^n
-\left(\sinh \case{V(z)-V_0}{2} \sinh \case{W-V_0}{2}\right) \lambda^{-n}
\nonumber \\
\quad
-\frac{\rmi k}{2} 
\biggl[
\cosh \case{W-V_0}{2}
\left(\rme^{[V(z)+V_0]/2} [\hbox{$-$}]_z^{x-nL}+\rme^{-[V(z)+V_0]/2}[\hbox{$+$}]_z^{x-nL}\right) \lambda^n
\nonumber \\
\quad \qquad \quad
+\sinh \case{W-V_0}{2}
\left(\rme^{[V(z)+V_0]/2} [\hbox{$-$}]_z^{x-nL}-\rme^{-[V(z)+V_0]/2}[\hbox{$+$}]_z^{x-nL}\right) \lambda^{-n}
\biggr]
\nonumber \\
\quad -\frac{\rmi k}{4L_0}\sinh\case{V(z)-W}{2}
\left(\,[\hbox{$-$}\hbox{$+$}]_{x-L}^{x}-\,[\hbox{$+$}\hbox{$-$}]_{x-L}^{x}\right)
(\lambda^n-\lambda^{-n})
 +O(k^2),
\end{eqnarray}
\begin{eqnarray}
\fl
\bar \beta(x,z;W;-k)=
\left(\sinh \case{V(z)-V_0}{2}\cosh \case{W-V_0}{2}\right)\lambda^n
-\left(\cosh \case{V(z)-V_0}{2} \sinh \case{W-V_0}{2}\right) \lambda^{-n}
\nonumber \\
\quad
-\frac{\rmi k}{2} 
\biggl[
\cosh \case{W-V_0}{2}
\left(\rme^{[V(z)+V_0]/2} [\hbox{$-$}]_z^{x-nL}-\rme^{-[V(z)+V_0]/2}[\hbox{$+$}]_z^{x-nL}\right) \lambda^n
\nonumber \\
\quad \qquad \quad
+\sinh \case{W-V_0}{2}
\left(\rme^{[V(z)+V_0]/2} [\hbox{$-$}]_z^{x-nL}+\rme^{-[V(z)+V_0]/2}[\hbox{$+$}]_z^{x-nL}\right) \lambda^{-n}
\biggr]
\nonumber \\
\quad -\frac{\rmi k}{4L_0}\cosh\case{V(z)-W}{2}
\left(\,[\hbox{$-$}\hbox{$+$}]_{x-L}^{x}-\,[\hbox{$+$}\hbox{$-$}]_{x-L}^{x}\right)
(\lambda^n-\lambda^{-n})
 +O(k^2).
\end{eqnarray}
\endnumparts
Details of the calculation are given in appendix~E.
From (\ref{2-5.20}) and (\ref{1-5.17}) we obtain
\refstepcounter{equation}
\label{1-5.18}
\addtocounter{equation}{-1}
\numparts
\begin{eqnarray}
\fl
\frac{\bar \tau^2(x,z;W;k)}{1-\bar R_l^2(x,z;W;k)}
&=\frac{1}{\cosh^2\frac{W-V_0}{2}}\,\frac{\gamma^n}{1-(c_0\gamma^n)^2}
\nonumber \\
\fl
& \qquad \times
\left(
1+ \rmi k\frac{1+c_0 \gamma^n}{1-c_0 \gamma^n}\rme^{-V_0}[\hbox{$+$}]_z^{x-nL}
+ \rmi k\frac{1-c_0 \gamma^n}{1+c_0 \gamma^n}\rme^{V_0}[\hbox{$-$}]_z^{x-nL}
+k B_1
\right)
\nonumber \\
\fl
&\quad +O(k^2),
\end{eqnarray}
\begin{eqnarray}
\fl
\frac{1+\bar R_l(x,z;W;k)}{1-\bar R_l(x,z;W;k)}
&=\rme^{-V(z)+V_0}\frac{1-c_0\gamma^n}{1+c_0\gamma^n}
\nonumber \\
\fl
& \qquad \times
\left(
1- \rmi k\frac{1+c_0 \gamma^n}{1-c_0 \gamma^n}\rme^{-V_0}[\hbox{$+$}]_z^{x-nL}
+ \rmi k\frac{1-c_0 \gamma^n}{1+c_0 \gamma^n}\rme^{V_0}[\hbox{$-$}]_z^{x-nL}
+ k B_2 \right)
\nonumber \\
\fl
&\quad +O(k^2),
\end{eqnarray}
\endnumparts
where $B_1$ and $B_2$ are quantities independent of $z$, and
\begin{equation}
\label{1-5.19}
c_0 \equiv -\tanh \frac{W-V_0}{2},  \qquad 
\gamma \equiv \frac{1}{\lambda^2}=1+2 \rmi k L_0 + O(k^2).
\end{equation}
As mentioned before, we cannot let $\gamma^n \simeq 1 + 2 n \rmi k L_0$ in (\ref{1-5.18}). We must take the sum over $n$ before using $ \gamma \simeq 1 + 2 \rmi k L_0$.
The $z$-independent terms $k B_1$ and $k B_2$ in (\ref{1-5.18}) derive from the terms involving $([\hbox{$-$}\hbox{$+$}]_{x-L}^{x}-\,[\hbox{$+$}\hbox{$-$}]_{x-L}^{x})$ in (\ref{1-5.17}). 
As we shall see, they are irrelevant to the final result.

We assume that $g(x,W)$ can be expanded in powers of $\rme^W$ as
\begin{equation}
\label{1-5.20}
g(x,W)=\sum_{j=0}^\infty g_j(x) \rme^{j W}.
\end{equation}
Condition (\ref{1-5.2}) requires that
\begin{equation}
\label{1-5.21}
\int_{x-L}^x g_j(z)\,\rmd z=0.
\end{equation}
For each term in (\ref{1-5.20}), expression (\ref{1-3.6}) gives
\begin{eqnarray}
\label{1-5.22}
\fl
(\mathcal{A}-2 \rmi k \mathcal{B})^{-1} g_j(x)\rme^{j W}
\nonumber \\
=
\int_{-\infty}^x \rmd z\,
\frac{\bar \tau^2(x,z;W;k)}{1-\bar R_l^2(x,z;W;k)}
\rme^{j V(z)}
\left[\frac{1+\bar R_l(x,z;W;k)}{1-\bar R_l(x,z;W;k)}\right]^j g_j(z).
\end{eqnarray}
We divide the integral as (\ref{1-5.3}), and shift the variable of integration by $z\to z-nL$. Thus, the right-hand side of (\ref{1-5.22}) is rewritten as
\begin{equation}
\label{1-5.23}
\int_{-\infty}^x C(z)\,\rmd z
=\sum_{n=0}^\infty\int_{x-(n+1)L}^{x-nL}C(z)\,\rmd z
=\sum_{n=0}^\infty\int_{x-L}^x C(z-nL)\,\rmd z,
\end{equation}
where $C(z)$ stands for the integrand in (\ref{1-5.22}). 
For each $n$ in the sum of (\ref{1-5.23}), we can use (\ref{1-5.18}).
Substituting (\ref{1-5.18}), and using $[\hbox{$+$}]_{z-nL}^{x-nL}=[\hbox{$+$}]_z^x$, $[\hbox{$-$}]_{z-nL}^{x-nL}=[\hbox{$-$}]_z^x$, we obtain
\begin{eqnarray}
\label{1-5.24}
\fl
&(\mathcal{A}-2 \rmi k \mathcal{B})^{-1}g_j(x)\rme^{j W}
\nonumber \\
\fl
& \qquad
=\frac{\rme^{j V_0}}{\cosh^2 \frac{W-V_0}{2}}\sum_{n=0}^\infty
\Biggl\{
\int_{x-L}^x
\frac{\gamma^n}{1-(c_0\gamma^n)^2}
\left(\frac{1-c_0 \gamma^n}{1+c_0 \gamma^n}\right)^j
\Biggr[
1 + \rmi k(1-j)\frac{1+c_0\gamma^n}{1-c_0\gamma^n}\rme^{-V_0}[\hbox{$+$}]_z^x
\nonumber \\
\fl
& \qquad \qquad \qquad \quad \qquad \quad
+ \rmi k(1+j)\frac{1-c_0\gamma^n}{1+c_0\gamma^n}\rme^{V_0}[\hbox{$-$}]_z^x
+k B_3
\Biggr] g_j(z)\,\rmd z 
+O(k^2)
\Biggr\},
\end{eqnarray}
where $B_3$ (which derives from $B_1$ and $B_2$ of (\ref{1-5.18})) is independent of $z$.
On the right-hand side, the term of order~$k^0$ vanishes on account of (\ref{1-5.21}). 
The part involving $B_3$ vanishes for the same reason. Therefore, (\ref{1-5.24}) becomes
\begin{eqnarray}
\label{1-5.25}
\fl
&(\mathcal{A}-2 \rmi k \mathcal{B})^{-1}g_j(x)\rme^{j W}
\nonumber \\
\fl
&\qquad
=\frac{\rmi k}{\cosh^2 \frac{W-V_0}{2}}
\sum_{n=0}^\infty
\Biggl\{
\frac{\gamma^n}{1-(c_0\gamma^n)^2}
\Biggl[
(j+1) \left(\frac{1-c_0\gamma^n}{1+c_0\gamma^n}\rme^{V_0}\right)^{j+1}
\int_{x-L}^x [\hbox{$-$}]_z^x\, g_j(z)\,\rmd z
\nonumber \\
\fl
& \qquad \qquad \qquad \qquad \quad 
- (j-1) \left(\frac{1-c_0\gamma^n}{1+c_0\gamma^n}\rme^{V_0} \right)^{j-1}
\int_{x-L}^x [\hbox{$+$}]_z^x\,  g_j(z)\,\rmd z
\Biggr]
+O(k)
\Biggr\}.
\end{eqnarray}

We have $\vert \gamma \vert<1$ as long as ${\rm Im}\,k>0$. (Recall that $R_r(k)\equiv \lim_{\epsilon \downarrow 0} R_r(k+\rmi \epsilon)$ for real~$k$.) 
For an arbitrary integer $m \geq 0$, we can calculate the infinite sum
\begin{equation}
\label{1-5.26}
\rmi k \sum_{n=0}^\infty \gamma^{n(1+m)}
=\frac{\rmi k}{1-\gamma^{1+m}}
=-\frac{1}{2(m+1)L_0}+O(k),
\end{equation}
where we have used
 $\gamma =1+2 \rmi k L_0 + O(k^2)
$. 
Hence, for any analytic function $h$,
\begin{equation}
\label{1-5.27}
\lim_{k\to 0} {\rmi k}\sum_{n=0}^\infty \gamma^n h(\gamma^n)
=-\frac{1}{2L_0}\int_0^1 h(x)\,\rmd x,
\end{equation}
as can be proved by Taylor expanding $h(x)$ and using (\ref{1-5.26}) for each term.
We can take the limit $k\to 0$ of (\ref{1-5.25}) by using (\ref{1-5.27}).
To calculate the integral, we use the formula
\begin{equation}
\label{1-5.28}
\fl
\int_0^1 \frac{1}{1-(c_0 x)^2}\left(\frac{1-c_0x}{1+c_0x}\right)^m\rmd x
=\frac{1}{2 m c_0}\left[1-\left(\frac{1-c_0}{1+c_0}\right)^m\right]
=\frac{\rme^{m(W-V_0)}-1}{2 m\tanh \frac{W-V_0}{2}}.
\end{equation}
From (\ref{1-5.25}), (\ref{1-5.27}) and (\ref{1-5.28}) we obtain
\begin{eqnarray}
\label{1-5.29}
\fl
\lim_{k \to 0} (\mathcal{A}-2 \rmi k\mathcal{B})^{-1} g_j(x) \rme^{j W}
\nonumber \\
=\frac{1}{2L_0 \sinh(W-V_0)}
\biggl\{
\left(\rme^{(j-1)W}-\rme^{(j-1)V_0}\right) \int_{x-L}^x[\hbox{$+$}]_z^x\,g_j(z)\,\rmd z
\nonumber \\
\qquad \qquad \qquad \qquad\qquad
-\left(\rme^{(j+1)W}-\rme^{(j+1)V_0}\right) \int_{x-L}^x[\hbox{$-$}]_z^x\,g_j(z)\,\rmd z
\biggr\}
\nonumber \\
=\frac{1}{2L_0 \sinh(W-V_0)}
\mathcal{D}\int_{x-L}^x\left(\rme^{-W}[\hbox{$+$}]_z^x-\rme^W [\hbox{$-$}]_z^x\right)
g_j(z) \rme^{j W} \rmd z.
\end{eqnarray}
Taking the sum over $j$, we arrive at the result\footnote{
%
%
Here we are assuming that $\mathcal{A}^{-1}\sum_j g_j \rme^{j W}=\sum_j \mathcal{A}^{-1}g_j \rme^{j W}$. If we can write $g(x,W)=\mathcal{A}h(x,W)$ with $h(x,W)=\sum_j h_j(x) \rme^{j W}$, this assumption is equivalent to $(\partial/\partial x) \sum_j h_j \rme^{j W}=\sum_j (\partial h_j/\partial x) \rme^{j W}$. 
It is not necessary to verify this assumption here, as it will be checked later (in section~6 and appendix~F) that (\ref{1-5.30}) is correct as long as $g(x,W)$ belongs to the range of $\mathcal{A}$ defined in the next section.
}
\begin{equation}
\label{1-5.30}
\fl
\mathcal{A}^{-1}g(x,W)=\frac{1}{2L_0 \sinh(W-V_0)}
\mathcal{D}\int_{x-L}^x\left(\rme^{-W}[\hbox{$+$}]_z^x-\rme^W [\hbox{$-$}]_z^x\right) 
g(z,W)\,\rmd z.
\end{equation}
Substituting $[\hbox{$+$}]_z^x=\int_z^x \rme^{V(z')}\, \rmd z'$ and $[\hbox{$-$}]_z^x=\int_z^x \rme^{-V(z')}\, \rmd z'$ into (\ref{1-5.30}) gives (\ref{1-4.3}).

The above derivation shows that $\lim_{k\to 0}(\mathcal{A}-2\rmi k  \mathcal{B})^{-1}g$ makes sense and is equal to the right-hand side of (\ref{1-4.3}) if both (\ref{1-5.1}) and (\ref{1-5.2}) are satisfied. If $g$ satisfies only (\ref{1-5.1}) and not (\ref{1-5.2}), then $(\mathcal{A}-2\rmi k  \mathcal{B})^{-1}g$ contains a term of order $k^{-1}$ (see (\ref{1-5.24}) and (\ref{1-5.27})), and so it is impossible to take the limit $k \to 0$. 

Since $(\partial/\partial x)\mathcal{A}^{-1}g=g$ (see section~6 and appendix~F), we can  express $\mathcal{A}^{-1} g$ as
\begin{equation}
\label{1-5.31}
\mathcal{A}^{-1}g(x,W)=\int_{x_0}^x g(z,W)\,\rmd z+c(x_0,W),
\end{equation}
where $x_0$ is an arbitrary number, and $c(x_0,W)$ is independent of $x$. 
Setting $x=x_0$ in (\ref{1-5.31}), we find $c(x_0,W)=(\mathcal{A}^{-1}g)(x_0,W)$. 
Equation ({\ref{1-5.31}) can also be directly derived by the method described in this section if we start with $\int_{-\infty}^x=\int_{x_0}^x + \sum \int_{x_0-(n+1)L}^{x_0-n L}$ instead of (\ref{1-5.3}).


\section{Domains of $\mathcal{A}$ and $\mathcal{A}^{-1}$}
Let $h$ and $g$ be functions such that $g(x,W)=\mathcal{A}h(x,W)$. 
We are assuming that $h(x,W)$ is piecewise continuously differentiable in $x$ and analytic in $W$ on the real axis, and also that 
\begin{equation}
\label{1-6.1}
h(x+L,W)=h(x,W).
\end{equation}
As shown in the last section, $\mathcal{A}^{-1}g(x,W)$ makes sense if (\ref{1-5.1}) and (\ref{1-5.2}) are satisfied. 
Equations (\ref{1-5.1}) and (\ref{1-5.2}) follow from (\ref{1-6.1}).
But this is not enough. For (\ref{1-4.3}) to give the proper inverse of $\mathcal{A}$, the domain of $\mathcal{A}$ must be further restricted.

As an inverse, $\mathcal{A}^{-1}$ is required to satisfy $\mathcal{A}\mathcal{A}^{-1} g=g$ and $\mathcal{A}^{-1}\mathcal{A} h=h$.
From the first line of (\ref{1-4.1}), it follows that
\refstepcounter{equation}
\label{1-6.2}
\addtocounter{equation}{-1}
\numparts
\begin{eqnarray}
\lim_{k \to 0}(\mathcal{A}-2\rmi k \mathcal{B})(\mathcal{A}-2\rmi k \mathcal{B})^{-1}
=\mathcal{A}\mathcal{A}^{-1}-2 \rmi \lim_{k\to 0}k \mathcal{B}\mathcal{A}^{-1},
\\
\lim_{k \to 0}(\mathcal{A}-2\rmi k \mathcal{B})^{-1}(\mathcal{A}-2\rmi k \mathcal{B})
=\mathcal{A}^{-1}\mathcal{A}-2 \rmi \lim_{k\to 0}k \mathcal{A}^{-1}\mathcal{B}.
\end{eqnarray}
\endnumparts
The left-hand sides of (\ref{1-6.2}) are identity operators. 
Therefore, $\mathcal{A}\mathcal{A}^{-1} g=g$ holds if $\mathcal{B}\mathcal{A}^{-1}g$ makes sense, and $\mathcal{A}^{-1}\mathcal{A} h=h$ holds if $\mathcal{A}^{-1}\mathcal{B}h$ makes sense. 
The former condition is always satisfied since $\mathcal{A}^{-1}g(x,W)$ is analytic in $W$. So, $\mathcal{A}\mathcal{A}^{-1}g=g$ holds without any further assumption on $g$.

  On the other hand, the condition for $\mathcal{A}^{-1}\mathcal{A} h=h$ imposes an additional restriction on $h$.
We already know that $\mathcal{A}^{-1} g$ makes sense if (\ref{1-5.2}) is satisfied. Replacing $g$ by $\mathcal{B}h$, we can see that $\mathcal{A}^{-1}\mathcal{B}h$ makes sense if
\begin{equation}
\label{1-6.3}
\int_{x-L}^x \mathcal{B}h(z,W)\,\rmd z=0.
\end{equation}
Namely, $\mathcal{A}^{-1}\mathcal{A} h=h$ holds if $h$ satisfies (\ref{1-6.3}).

In appendix~F, it is verified by direct calculation that the operator $\mathcal{A}^{-1}$ given by (\ref{1-4.3}) (or (\ref{1-5.30}))  satisfies $\mathcal{A}^{-1}\mathcal{A} h=h$ and $\mathcal{A}\mathcal{A}^{-1} g=g$, provided that $h$ and $g$ satisfy the conditions specified above. Hence, we know that (\ref{1-4.3}) is indeed the correct expression for the inverse of $\mathcal{A}$.

The conclusion of this section is as follows.
We take the domain of the differential operator $\mathcal{A}=\partial/\partial x$ to be the set of functions $h(x,W)$ which satisfy (\ref{1-6.1}) and (\ref{1-6.3}), and which are piecewise continuously differentiable in $x$ and analytic in $W$ on the real axis. Then (\ref{1-4.3}) gives the inverse of $\mathcal{A}$, satisfying $\mathcal{A}^{-1}\mathcal{A}h=h$. 
The domain of $\mathcal{A}^{-1}$ is the range of $\mathcal{A}$. It consists of functions $g(x,W)$ which satisfy (\ref{1-5.1}) and (\ref{1-5.2}), and which are analytic in $W$ and continuous in $x$ except possibly for some finite jumps and delta function singularities.

Let us note that $\bar R_r(x,-\infty;W;k)+\xi(x,W)$, which appears on the left-hand side of (\ref{1-3.1}), belongs to the domain of $\mathcal{A}$ defined above. 
It is obvious that $\bar R_r+\xi$ satisfies (\ref{1-6.1}).
To see that (\ref{1-6.3}) is satisfied, we rewrite (\ref{1-3.1}) as
\begin{equation}
\mathcal{B} (\bar R_r +\xi)
=\frac{1}{2 \rmi k}
\left[
\frac{\partial}{\partial x}(\bar R_r +\xi) -(1-\xi^2) f
\right]
=\frac{1}{2 \rmi k} \frac{\partial}{\partial x}\bar R_r,
\end{equation}
where we have used $(1-\xi^2)f=(\partial /\partial x)\xi$ (equation (3.14)). 
Therefore,
\begin{equation}
\label{1-7.1}
\fl
\int_{x-L}^x \rmd z\, \mathcal{B}\left[\bar R_r(z,-\infty;W;k)+\xi(z,W)\right]
=\frac{1}{2 \rmi k}\int_{x-L}^x \rmd z\,
\frac{\partial}{\partial z}\bar R_r(z,-\infty;W;k).
\end{equation}
The right-hand side vanishes because $\bar R_r(x,-\infty;W;k)$ is periodic in $x$.
Thus $\bar R_r+\xi$ satisfies condition (\ref{1-6.3}).
It is not difficult to show, by using (\ref{1-3.1}) and (\ref{1-2.12b}), that $\bar R_r+\xi$ is piecewise continuously differentiable in $x$ and analytic in $W$ on the real axis.


\section{Validity of the expansion}
Consistency between (\ref{1-6.3}) and $g=\mathcal{A}h$ requires that
\begin{equation}
\label{1-6.4}
\int_{x-L}^x \mathcal{B}\mathcal{A}^{-1}g(z,W)\,\rmd z=0.
\end{equation}
As shown in appendix~G, equation (\ref{1-6.4}) holds if $g$ satisfies (\ref{1-5.2}). This means that $\mathcal{B}\mathcal{A}^{-1}g$ is in the domain of $\mathcal{A}^{-1}$ if $g$ is in the domain of $\mathcal{A}^{-1}$. Recursively, it follows that $(\mathcal{B}\mathcal{A}^{-1})^n g$ belongs to the domain of $\mathcal{A}^{-1}$ if $g$ is in the domain of $\mathcal{A}^{-1}$. In other words, $\mathcal{A}^{-1}(\mathcal{B}\mathcal{A}^{-1})^n g$ makes sense for any $n$ if $\mathcal{A}^{-1}g$ makes sense. 

New we can see that equations (\ref{1-3.11}), (\ref{1-3.12}) and (\ref{1-3.13}) are valid for periodic potentials if (\ref{1-3.3}) is replaced by (\ref{1-4.3}).
Since (\ref{1-3.12b}) can be written as 
\begin{equation}
\bar r_n=2^n\mathcal{A}^{-1}(\mathcal{B} \mathcal{A}^{-1})^n (1-\xi^2) f
\qquad (n\geq 1),
\end{equation}
it follows from the above argument that $\bar r_n$ makes sense for any $n$ if $(1-\xi^2) f$ belongs to the domain of $\mathcal{A}^{-1}$. 
From (\ref{1-3.14}) it is obvious that $(1-\xi^2)f$ satisfies conditions (\ref{1-5.1}) and (\ref{1-5.2}). Therefore $(1-\xi^2)f$ is in the domain of $\mathcal{A}^{-1}$, and hence $\bar r_0$ and all $\bar r_n$ make sense. 
It is easy to show that $(\mathcal{A}-2 \rmi k \mathcal{B})^{-1}g$ makes sense if $\mathcal{A}^{-1}g$ makes sense. So the right-hand side of (\ref{1-3.13}) makes sense, too. 

Thus, expansion (\ref{1-3.11}) is justified for any $N$. The coefficients of the expansion $\bar r_n$ given by (\ref{1-3.12}), as well as the remainder term $\bar \rho_N$, all make sense and are finite.

The behavior of the remainder term as $k\to 0$ is simple in the periodic case. From (\ref{1-3.13}) and (\ref{1-4.1}) we have $\lim_{k\to 0}\bar \rho_N/(\rmi k)^{N+1}=\mathcal{A}^{-1}\mathcal{A}\, \bar r_{N+1}=\bar r_{N+1}$. Therefore, $\bar \rho_N=O(k^{N+1})$ as $k\to 0$, since $\bar r_n$ is finite for any $n$. Unlike in the non-periodic cases \cite{low}, no subtleties arise when the potential is periodic. Expansion (\ref{1-3.11}) is asymptotic for any $N$. Moreover, the infinite series $\sum_{n=0}^\infty (\rmi k)^n \bar r_n$ is convergent for sufficiently small $\vert k \vert$. An explanation about the convergence of this series is given in appendix~B.


\section{Expressions for $\boldsymbol{\bar r}_n$}
Now let us calculate the coefficients $\bar r_0, \bar r_1, \bar r_2,\ldots$ of (\ref{1-3.11}).
First, $\bar r_0$ is given by (\ref{1-3.12a}). 
Substituting (\ref{1-3.14}) into (\ref{1-4.3}), and using 
$
\int_{x-L}^x\rme^{\pm V(z')}\,\rmd z'=L_0 \,\rme^{\pm V_0}
$, we obtain
\begin{eqnarray}
\label{1-8.1} 
\fl
\mathcal{A}^{-1}(1-\xi^2)f
= \frac{1}{L_0 \sinh(W-V_0)} \mathcal{D}\int_{x-L}^x\rmd z'\int_{x-L}^{z'} \rmd z 
\sinh[V(z')-W]
\frac{\partial}{\partial z} \tanh\case{W-V(z)}{2}
\nonumber \\
\fl \qquad 
= \frac{1}{L_0 \sinh(W-V_0)} 
\mathcal{D}\int_{x-L}^x \rmd z' \sinh[V(z')-W]
\left(
\tanh\case{W-V(z')}{2}-\tanh\case{W-V(x)}{2}
\right) 
\nonumber \\
\fl  \qquad
= \frac{1}{L_0 \sinh(W-V_0)} 
\mathcal{D}\left[L-L_0\cosh(W-V_0)+L_0 \sinh(W-V_0)\,\xi(x,W)\right]
\nonumber \\
\fl \qquad
=\frac{1}{\sinh(W-V_0)} 
\left[
1-\cosh(W-V_0) + \sinh(W-V_0)\,\xi(x,W)
\right]
\nonumber \\
\fl \qquad 
=-\tanh \frac{W-V_0}{2} +\xi,
\end{eqnarray}
and hence 
\begin{equation}
\label{1-8.2}
\bar r_0=-\tanh \frac{W-V_0}{2}.
\end{equation}
This is to be compared with (\ref{1-3.16}).

Next, we calculate $\bar r_1=2 \mathcal{A}^{-1}\mathcal{B}(\bar r_0 +\xi)$.
We can see that
\begin{eqnarray}
\label{1-8.3}
\fl
\mathcal{B}\left(\bar r_0 + \xi\right)
&=
\frac{\partial}{\partial W}
\left\{\sinh[W-V(x)]\left(\tanh\frac{V_0-W}{2}+\tanh\frac{W-V(x)}{2}\right)\right\}
\nonumber \\
\fl
&=
\frac{\partial}{\partial W}
\left\{\sinh[V_0-V(x)]\left(\tanh\frac{W-V_0}{2}+\tanh\frac{V_0-V(x)}{2}\right)\right\}
\nonumber \\
\fl
&=\sinh[V_0-V(x)]\,\frac{1}{2\cosh^2 \frac{W-V_0}{2}}.
\end{eqnarray}
Substituting this into (\ref{1-4.3}), and multiplying by 2, we have
\begin{eqnarray}
\label{1-8.4}
\fl
\bar r_1 
&=\frac{1}{L_0 \sinh(W-V_0)}
\nonumber \\
\fl
& \qquad \times
\,\mathcal{D}\frac{1}{\cosh^2 \frac{W-V_0}{2}}\int \!\!\!\int_{x-L\leq z\leq z'\leq x} \rmd z\, \rmd z'\,\sinh[V(z')-W] \sinh[V_0-V(z)].
\end{eqnarray}
The integrals on the right-hand side can be expressed as
\begin{eqnarray}
\label{1-8.5}
\fl
\int \!\!\!\int_{x-L\leq z\leq z'\leq x} \rmd z\, \rmd z'\,\sinh[V(z')-W] \sinh[V_0-V(z)]
\nonumber \\
\fl
\qquad =
\frac{1}{4}
\left(
\rme^{-W+V_0}[\hbox{$-$$+$}]_{x-L}^x+\rme^{W-V_0}[\hbox{$+$$-$}]_{x-L}^x
-\rme^{-W-V_0}[\hbox{$+$$+$}]_{x-L}^x-\rme^{W+V_0}[\hbox{$-$$-$}]_{x-L}^x
\right).
\end{eqnarray}
Substituting (\ref{1-8.5}) into (\ref{1-8.4}), and using definition (\ref{1-2.23}), we find
\begin{eqnarray}
\label{1-8.6}
\fl
\bar r_1=\frac{1}{4 L_0 \sinh(W-V_0)}
&\Biggl\{
\left(\frac{\rme^{-W+V_0}}{\cosh^2\frac{W-V_0}{2}}-1 \right)
\left(\,[\hbox{$-$$+$}]_{x-L}^x- \rme^{-2V_0}[\hbox{$+$$+$}]_{x-L}^x\right)
\nonumber \\
\fl
& \qquad +
\left(\frac{\rme^{W-V_0}}{\cosh^2\frac{W-V_0}{2}}-1 \right)
\left(\,[\hbox{$+$$-$}]_{x-L}^x- \rme^{2V_0}[\hbox{$-$$-$}]_{x-L}^x \right)
\Biggr\}.
\end{eqnarray}
Using (\ref{1-2.18}) and (\ref{1-2.19}), we can reduce (\ref{1-8.6}) to the simple form
\begin{equation}
\label{1-8.7}
\bar r_1=\frac{1}{4 L_0 \cosh^2 \frac{W-V_0}{2}}
\left(
\,[\hbox{$+$$-$}]_{x-L}^x-\,[\hbox{$-$$+$}]_{x-L}^x
\right).
\end{equation}

For $n\geq 2$, we can calculate $\bar r_n$ by successively applying $\mathcal{L}$ to $\bar r_1$ as
$
\bar r_n=\mathcal{L}^{n-1}\bar r_1
$.
Let us write the operator $\mathcal{B}$ as
\begin{equation}
\label{1-8.8}
\mathcal{B}=\frac{1}{2}\sum_{\sigma=\pm 1} \rme^{-\sigma V(x)}\hat \mathcal{J}_\sigma,
\qquad
\hat \mathcal{J}_\sigma \equiv \rme^{\sigma W}\left(1+\sigma \frac{\partial}{\partial W}\right).
\end{equation}
Substituting
$
\sinh[V(z')-W]=\frac{1}{2}\sum_{\sigma'=\pm 1}\sigma' \rme^{\sigma' V(z')}\rme^{-\sigma' W}
$
into (\ref{1-4.3}), and using (\ref{1-8.8}), we can write $\mathcal{L}$ in the form
\begin{equation}
\label{1-8.9}
\mathcal{L}=2\mathcal{A}^{-1}\mathcal{B}
=\frac{1}{2L_0}
\sum_{\sigma=\pm 1}\sum_{\sigma'=\pm 1}\mathcal{I}_{\sigma,\sigma'}\mathcal{K}_{-\sigma,-\sigma'},
\end{equation}
where we have defined the operators $\mathcal{I}_{\sigma,\sigma'}$ and $\mathcal{K}_{\sigma,\sigma'}$ as
\begin{equation}
\label{1-8.10}
\mathcal{I}_{\sigma,\sigma'}g(x,W) \equiv \int \!\!\! \int_{x-L\leq z\leq z'\leq x} \rmd z \,\rmd z'
\,\rme^{\sigma V(z) +\sigma' V(z')} g(z,W),
\end{equation}
\begin{eqnarray}
\label{1-8.11}
\mathcal{K}_{\sigma,\sigma'}g(x,W)&\equiv
\frac{\sigma'}{\sinh(V_0-W)}\mathcal{D}\, \rme^{\sigma' W} \hat \mathcal{J}_\sigma g(x,W)
\nonumber \\
&=\frac{1}{\sinh(V_0-W)}\int_{V_0}^W \rmd W
\,\hat \mathcal{J}_{\sigma'} \hat \mathcal{J}_\sigma \,g(x,W).
\end{eqnarray}
The expression $\bar r_2=\mathcal{L}\,\bar r_1$ with (\ref{1-8.7}) and (\ref{1-8.9}) reads
\begin{equation}
\label{1-8.12}
\fl
\bar r_2=\frac{1}{8L_0^2}\sum_{\sigma=\pm 1}\sum_{\sigma'=\pm 1}
\left(\mathcal{K}_{-\sigma,-\sigma'} \,\frac{1}{\cosh^2\frac{W-V_0}{2}}
\right)
\mathcal{I}_{\sigma,\sigma'}
\left(
\,[\hbox{$+$$-$}]_{x-L}^x-\,[\hbox{$-$$+$}]_{x-L}^x
\right).
\end{equation}
Note that
\begin{eqnarray}
\label{1-8.13}
[\sigma_1,\sigma_2]_{z-L}^z 
&= [\sigma_1,\sigma_2]_{x-L}^z +[\sigma_1]_{z-L}^{x-L}\,[\sigma_2]_{x-L}^z+[\sigma_1,\sigma_2]_{z-L}^{x-L}
\nonumber \\
&= [\sigma_1,\sigma_2]_{x-L}^z +[\sigma_2]_{x-L}^z\,[\sigma_1]_z^x+[\sigma_1,\sigma_2]_z^x.
\end{eqnarray}
Therefore,
\begin{eqnarray}
\label{1-8.14}
\fl
\mathcal{I}_{\sigma,\sigma'}[\sigma_1,\sigma_2]_{x-L}^x 
&=
 \int \!\!\! \int_{x-L\leq z\leq z'\leq x} \rmd z \,\rmd z'
\,\rme^{\sigma V(z) +\sigma' V(z')} [\sigma_1,\sigma_2]_{z-L}^z
\nonumber \\
&=[\sigma_1,\sigma_2,\sigma,\sigma']_{x-L}^x+[\sigma_2,\sigma,\sigma_1,\sigma']_{x-L}^x+[\sigma_2,\sigma,\sigma',\sigma_1]_{x-L}^x
\nonumber \\
& \qquad +[\sigma,\sigma_1,\sigma_2,\sigma']_{x-L}^x+[\sigma,\sigma_1,\sigma',\sigma_2]_{x-L}^x+[\sigma,\sigma',\sigma_1,\sigma_2]_{x-L}^x.
\end{eqnarray}
This gives explicitly
\refstepcounter{equation}
\label{1-8.15}
\addtocounter{equation}{-1}
\numparts
\begin{eqnarray}
\fl
\mathcal{I}_{++}
\left(
\,[\hbox{$+$$-$}]_{x-L}^x-\,[\hbox{$-$$+$}]_{x-L}^x
\right)
\nonumber \\
=[\hbox{$-$$+$$+$$+$}]_{x-L}^x+[\hbox{$+$$+$$+$$-$}]_{x-L}^x-[\hbox{$+$$-$$+$$+$}]_{x-L}^x-[\hbox{$+$$+$$-$$+$}]_{x-L}^x
\nonumber \\
=\case{1}{6}P^3M-P\,[\hbox{$+$$-$$+$}]_{x-L}^x,
\end{eqnarray}
\begin{eqnarray}
\fl
\mathcal{I}_{--}
\left(
\,[\hbox{$+$$-$}]_{x-L}^x-\,[\hbox{$-$$+$}]_{x-L}^x
\right)
\nonumber \\
=-[\hbox{$+$$-$$-$$-$}]_{x-L}^x-[\hbox{$-$$-$$-$$+$}]_{x-L}^x+[\hbox{$-$$-$$+$$-$}]_{x-L}^x+[\hbox{$-$$+$$-$$-$}]_{x-L}^x
\nonumber \\
=-\case{1}{6}M^3P+M\,[\hbox{$-$$+$$-$}]_{x-L}^x,
\end{eqnarray}
\begin{eqnarray}
\fl
\mathcal{I}_{+-}
\left(
\,[\hbox{$+$$-$}]_{x-L}^x-\,[\hbox{$-$$+$}]_{x-L}^x
\right)
&=[\hbox{$-$$+$$-$$+$}]_{x-L}^x+[\hbox{$+$$-$$+$$-$}]_{x-L}^x-2\,[\hbox{$+$$-$$-$$+$}]_{x-L}^x
\nonumber \\
&=2Q-M \,[\hbox{$+$$-$$+$}]_{x-L}^x,
\end{eqnarray}
\begin{eqnarray}
\fl
\mathcal{I}_{-+}
\left(
\,[\hbox{$+$$-$}]_{x-L}^x-\,[\hbox{$-$$+$}]_{x-L}^x
\right)
&=-[\hbox{$-$$+$$-$$+$}]_{x-L}^x-[\hbox{$+$$-$$+$$-$}]_{x-L}^x+2\,[\hbox{$-$$+$$+$$-$}]_{x-L}^x
\nonumber \\
&=-2 Q+P \,[\hbox{$-$$+$$-$}]_{x-L}^x,
\end{eqnarray}
\endnumparts
where we have used (\ref{1-2.16}) and defined
\begin{equation}
\label{1-8.16}
Q \equiv [\hbox{$-$$+$$-$$+$}]_{x-L}^x+[\hbox{$+$$-$$+$$-$}]_{x-L}^x.
\end{equation}
This $Q$ is independent of $x$, as can be seen by differentiating the right-hand side with respect to $x$ and using (\ref{1-2.20}).
The expression in large parentheses in (\ref{1-8.12}) can be calculated as
\begin{equation}
\label{1-8.17}
\mathcal{K}_{-\sigma,-\sigma'} \,\frac{1}{\cosh^2\frac{W-V_0}{2}}
=\frac{\sigma'\rme^{-(\sigma+\sigma')V_0}}{\sinh(W-V_0)}
\left(
\frac{\rme^{-[(\sigma/2)+\sigma'](W-V_0)}}{\cosh^3\frac{W-V_0}{2}} 
-1
\right).
\end{equation}
Substituting (\ref{1-8.15}) and (\ref{1-8.17}) into (\ref{1-8.12}), we obtain
\begin{eqnarray}
\label{1-8.18}
\fl 
\bar r_2 =\frac{1}{4L_0\cosh^3\frac{W-V_0}{2}}
\Biggl\{&
\rme^{-(W+V_0)/2}\,[\hbox{$+$$-$$+$}]_{x-L}^x
-\rme^{(W+V_0)/2}\,[\hbox{$-$$+$$-$}]_{x-L}^x 
\nonumber \\
& \qquad \qquad \qquad \qquad +\frac{1}{L_0}\left(
\frac{L_0^4}{4} +Q \right)\sinh\frac{W-V_0}{2}
\Biggr\}.
\end{eqnarray}

The coefficient $\bar r_n$ for $n\geq 3$ can be obtained in the same way.
The expression for general $n$ is
\begin{eqnarray}
\label{1-8.19}
\fl
\bar r_n=\frac{1}{2^{n+1}L_0^n}\sum_{\{\sigma_i,\sigma'_i\}}
&\left(
\mathcal{K}_{-\sigma_{n-1},-\sigma'_{n-1}}\cdots \mathcal{K}_{-\sigma_2,-\sigma'_2}\mathcal{K}_{-\sigma_1,-\sigma'_1} \,\frac{1}{\cosh^2\frac{W-V_0}{2}}
\right)
\nonumber \\
& \qquad \times\mathcal{I}_{\sigma_{n-1},\sigma'_{n-1}}\cdots \mathcal{I}_{\sigma_2,\sigma'_2}\mathcal{I}_{\sigma_1,\sigma'_1}
\left(
\,[\hbox{$+$$-$}]_{x-L}^x-\,[\hbox{$-$$+$}]_{x-L}^x
\right),
\end{eqnarray}
where the sum is over $\sigma_i=\pm 1$ and $\sigma_i'=\pm 1$ for $1\leq i \leq n-1$.


\section{The expansion of $\boldsymbol{S}$}
It was shown in \cite{low} that the expansion of $S_r$ takes the form
\begin{equation}
\label{1-9.1}
S_r(x,k)-\frac{1}{2}=a_0(x) + \rmi k a_1(x) +(\rmi k)^2a_2(x)+(\rmi k)^3a_3(x)+\cdots,
\end{equation}
where
\refstepcounter{equation}
\label{1-9.2}
\addtocounter{equation}{-1}
\numparts
\begin{equation}
a_0=\frac{1}{4}\lim_{W \to -\infty}\rme^{-W+V(x)}(\bar r_0-1),
\end{equation}
\begin{equation}
a_n=\frac{1}{4}\lim_{W \to -\infty}\rme^{-W+V(x)}\,\bar r_n
\qquad (n\geq 1)
\end{equation}
\endnumparts
(see equations~(4.4) of \cite{low}).
Substituting (\ref{1-8.2}), (\ref{1-8.7}) and (\ref{1-8.18}) into (\ref{1-9.2}) gives
\refstepcounter{equation}
\label{1-9.3}
\addtocounter{equation}{-1}
\numparts
\begin{equation}
a_0(x)=-\frac{1}{2} \rme^{V(x)-V_0},
\end{equation}
\begin{equation}
a_1(x)=\frac{1}{4 L_0} \rme^{V(x)-V_0}
\left(\,[\hbox{$+$$-$}]_{x-L}^x -[\hbox{$-$$+$}]_{x-L}^x\right),
\end{equation}
\begin{equation}
a_2(x)=\frac{1}{2 L_0} \rme^{V(x)-V_0}
\left\{
 \rme^{-V_0}\,[\hbox{$+$$-$$+$}]_{x-L}^x 
 -\frac{1}{2 L_0}\left(\frac{L_0^4}{4}+Q\right)
\right\}.
\end{equation}
\endnumparts

When the potential is periodic, the functions $S_r$ and $S_l$ are simply related by
\begin{equation}
\label{1-9.4}
S_r(x,k)=S_l(x,-k).
\end{equation}
(See appendix~D. This is a special feature of the periodic case.) Therefore,
\begin{equation}
\label{1-9.5}
S(x,k)-1=s_0(x)+ (\rmi k)^2 s_2(x)+(\rmi k)^4 s_4(x)+\cdots,
\end{equation}
where
\begin{equation}
\label{1-9.6}
s_n(x)=2 a_n(x) \quad (\hbox{$n$ even}), \qquad s_n(x)=0 \quad (\hbox{$n$ odd}).
\end{equation}
The first two nonvanishing coefficients are
\refstepcounter{equation}
\label{1-9.7}
\addtocounter{equation}{-1}
\numparts
\begin{equation}
s_0(x)=-\rme^{V(x)-V_0},
\end{equation}
\begin{equation}
s_2(x)=\frac{\rme^{V(x)-V_0}}{L_0}
\left\{\rme^{-V_0}\,[\hbox{$+$$-$$+$}]_{x-L}^x
 -\frac{1}{2L_0}\left(\frac{L_0^4}{4}+Q\right)\right\}.
\end{equation}
\endnumparts


\section{The expansion of the Green function}
The small-$k$ expansion of the Green function is obtained by substituting (\ref{1-9.5}) into (\ref{1-2.10}). 
The expansion has the form
\begin{equation}
\label{1-10.1}
G_{\rm S}(x,y;k)=(\rmi k)^{-1}g_{-1}+g_0 + \rmi k g_1 +(\rmi k)^2g_2 + \cdots,
\end{equation}
and the coefficients can be expressed in terms of $s_0, s_1, s_1,\ldots$ as
\refstepcounter{equation}
\label{1-10.2}
\addtocounter{equation}{-1}
\numparts
\begin{equation}
g_{-1}(x,y)=\frac{1}{2 \sqrt{s_0(x)s_0(y)}},
\end{equation}
\begin{equation}
g_0(x,y)=q_1(x,y) g_{-1}(x,y),
\end{equation}
\begin{equation}
g_1(x,y)=\frac{1}{2}\left\{
[q_1(x,y)]^2-\frac{s_2(x)}{s_0(x)}-\frac{s_2(y)}{s_0(y)}
\right\}g_{-1}(x,y),
\end{equation}
\begin{equation}
\fl
g_2(x,y)=\left\{
\frac{1}{6}[q_1(x,y)]^3+q_3(x,y)-\frac{1}{2}q_1(x,y)\left[\frac{s_2(x)}{s_0(x)}+\frac{s_2(y)}{s_0(y)}\right]
\right\}g_{-1}(x,y),
\end{equation}
\endnumparts
where we have defined
\begin{equation}
\label{1-10.3}
q_n(x,y) \equiv -\int_y^x s_{n-1}(z)\,\rmd z.
\end{equation}
Substituting (\ref{1-9.7}) into (\ref{1-10.2}), and also using $q_1(x,y)=\rme^{-V_0}[\hbox{$+$}]_y^x$, we obtain
\refstepcounter{equation}
\label{1-10.4}
\addtocounter{equation}{-1}
\numparts 
\begin{equation}
g_{-1}(x,y)=\frac{\rme^{-[V(x)+V(y)]/2}}{2}\,\rme^{V_0},
\end{equation}
\begin{equation}
g_0(x,y)=\frac{\rme^{-[V(x)+V(y)]/2}}{2} \,[\hbox{$+$}]_y^x,
\end{equation}
\begin{eqnarray}
\fl
g_1(x,y)=\frac{\rme^{-[V(x)+V(y)]/2}}{4 L_0}
\Biggl\{&
\,[\hbox{$+$$-$$+$}]_{x-L}^x+[\hbox{$+$$-$$+$}]_{y-L}^y
\nonumber \\
\fl
& \qquad
+L_0\,\rme^{-V_0} \left([\hbox{$+$}]_y^x\right)^2
-\frac{\rme^{V_0}}{L_0}\left(\frac{L_0^4}{4}+Q\right)
\Biggr\},
\end{eqnarray}
\begin{eqnarray}
\fl
g_2(x,y)=\rme^{-V_0}\,[\hbox{$+$}]_y^x\,g_1(x,y)
-\left\{
\frac{\rme^{-3V_0}}{3}\left([\hbox{$+$}]_y^x\right)^3+\int_y^x s_2(z)\,\rmd z
\right\}g_{-1}(x,y).
\end{eqnarray}
\endnumparts
The series (\ref{1-10.1}) is convergent for sufficiently small $\vert k \vert$ (see the comment at the end of section~7 and appendix~B).


\section{Example}
As an example, let us consider the periodic square potential \cite{vollmer}
\begin{equation}
\label{1-11.1}
V(x)=
\cases{
0 & $(0<x<a)$ \\
C & $(a<x<L)$
}, 
\qquad
V(x+L)=V(x).
\end{equation}
For $0<x<a$, we have the exact expressions of $\alpha$ and $\beta$ as \cite{theory}
\refstepcounter{equation}
\label{1-11.2}
\addtocounter{equation}{-1}
\numparts
\begin{equation}
\alpha(x,x-L;k)=\frac{1}{1-A^2}\rme^{-\rmi k L}\left(1-A^2 \rme^{2\rmi k b}\right),
\end{equation}
\begin{equation}
\beta(x,x-L;k)=\frac{A }{1-A^2}\rme^{2\rmi k x} \rme^{-\rmi k L}\left(\rme^{2\rmi k b}-1\right),
\end{equation}
\endnumparts
where
\begin{equation}
\label{1-11.3}
A \equiv -\tanh (C/2), \qquad b \equiv L-a.
\end{equation}
From (\ref{1-11.2}) we obtain $R_r$, $R_l$ and $S$ by using (\ref{1-a.6}) and (\ref{1-a.8}) of appendix~D.
The expressions for $a<x<L$ have similar forms. 
The exact expression of the Green function is obtained by substituting (\ref{1-a.8}) into (\ref{1-2.10}), 
or, more simply, by using equations~(3.6) of \cite{expressions}. 
The resulting expression of $G_{\rm S}$ for $0<y\leq x <a$ is
\begin{equation}
\label{1-11.4}
\fl
G_{\rm S}(x,y;k)=
\frac{
\left[2 A \rme^{2 \rmi k x} \rme^{-\rmi k (L-b)} \sin k b- K(k)\right]
\left[2 A \rme^{-2 \rmi k y} \rme^{\rmi k (L-b)} \sin k b- K(k)\right]
}
{
2 \rmi k \rme^{\rmi k (x-y)}
\left\{4 A^2 \sin^2 k b -[K(k)]^2 \right\}
},
\end{equation}
where
\begin{equation}
\label{1-11.5}
K(k) \equiv \sin k L-A^2 \sin[k(L-2 b)]-(1-A^2) \left\{1-[Y(k)]^2\right\}^{1/2},
\end{equation}
\begin{equation}
\label{1-11.6}
Y(k)=\frac{1}{1-A^2}\left\{ \cos k L - A^2 \cos[k(L-2b)]\right\}
\end{equation}
(see (\ref{1-5.7}) for the definition of $Y$). 

It is characteristic of periodic systems that the energy spectrum has a band structure. Let us suppose that $k$ is a real number. The energy lies in a band if $[Y(k)]^2 <1$, and in a gap if $[Y(k)]^2>1$ (see appendix~D).
The band structure can be seen in the graph of $G_{\rm S}(k)$ (figure~1). The low-energy expansion can be used for calculating the Green function in the lowest band.
%
%
%
\begin{figure}
\hspace{1cm}
\includegraphics[scale=0.75]{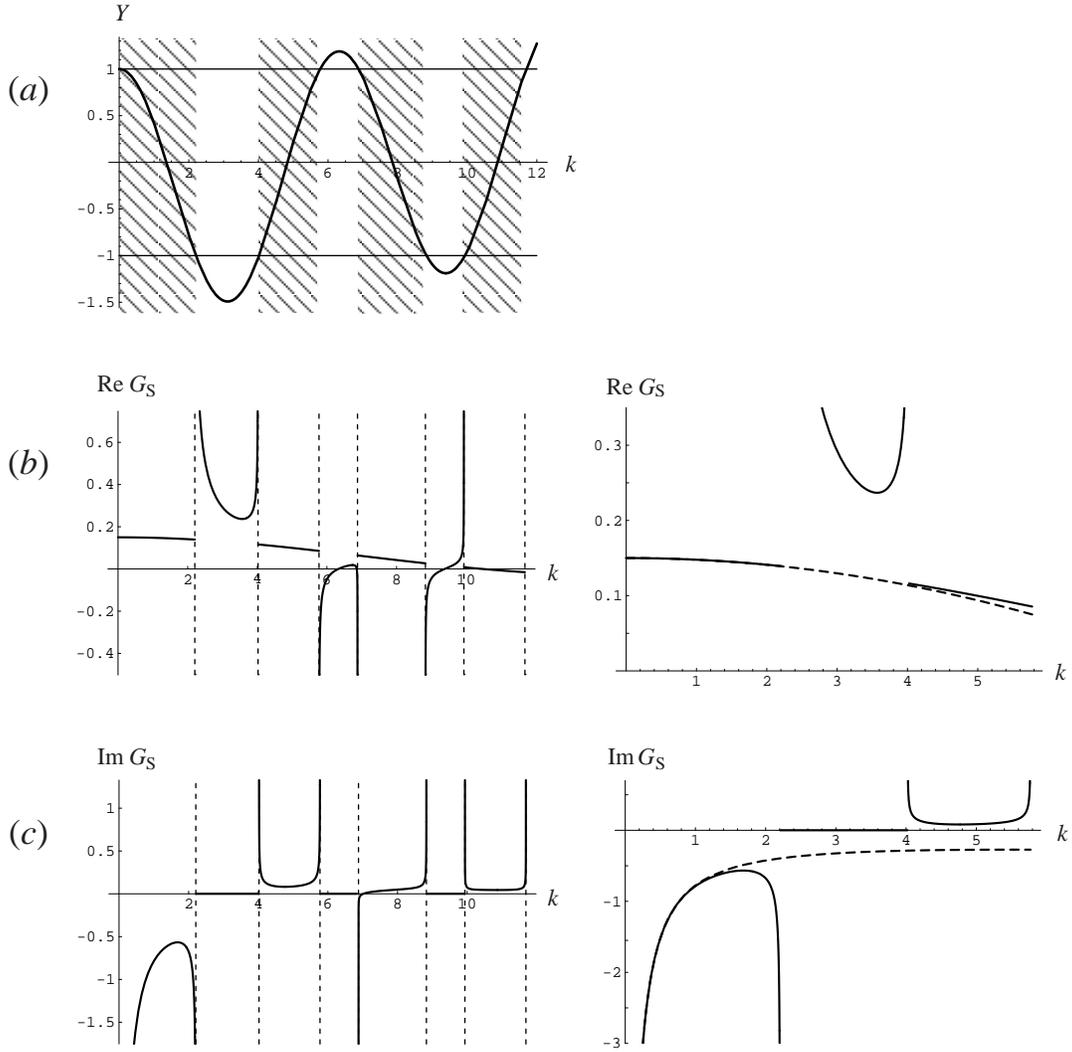}
\caption{
Graphs of (a) $Y(k)$, (b) ${\rm Re\,}G_{\rm S}(x,y;k)$ and (c) ${\rm Im\,}G_{\rm S}(x,y;k)$ for the potential (\ref{1-11.1}), plotted as functions of real $k$, with $C=1$, $L=1$, $a=0.6$, $x=0.4$ and $y=0.1$. 
In (a), the shaded and unshaded areas represent the bands ($\vert Y \vert<1$) and the gaps ($\vert Y \vert>1$), respectively. 
In (b) and (c), the solid lines show the exact value of $G_{\rm S}$ (equation (\ref{1-11.4})). 
The imaginary part of $G_{\rm S}$ is identically zero in the gaps. The graphs on the right are close-up plots, where the dashed lines show the result of the low-energy expansion up to order~$k^2$. The dashed curves almost overlap with the solid curves in the lowest band except for the imaginary part near the band edge.}
\end{figure}

In this example, $M=a+b \,\rme^{-C}$ and $P=a+b\,\rme^C$, as can be easily seen. Hence,
\begin{equation}
\label{1-11.7}
L_0=\sqrt{(a+b \,\rme^{-C})(a+b\,\rme^C)},
\qquad
\rme^{V_0}=\sqrt{ \frac{a+b\,\rme^C}{a+b \,\rme^{-C}}}.
\end{equation}
We assume that $0<x<a$.
The integrals appearing in the expansion of $\bar R_r$ or $S$ can be calculated, for example, as
\begin{eqnarray}
\label{1-11.8}
\fl
[\hbox{$+$}\hbox{$-$}]_{x-L}^x
&=[\hbox{$+$}\hbox{$-$}]_0^x
+[\hbox{$+$}\hbox{$-$}]_{-b}^0
+[\hbox{$+$}\hbox{$-$}]_{x-L}^{-b}
+[\hbox{$+$}]_{x-L}^{-b}[\hbox{$-$}]_{-b}^0
+[\hbox{$+$}]_{x-L}^{-b}[\hbox{$-$}]_0^x
+[\hbox{$+$}]_{-b}^0[\hbox{$-$}]_0^x
\nonumber \\
\fl
&=\frac{1}{2}\left[x^2+b^2+(a-x)^2\right]+\rme^{-C}(a-x)b +(a-x) x+ \rme^C b x
\nonumber \\
\fl
&=\frac{1}{2}(a^2+b^2)+\rme^{-C}b(a-x)+ \rme^C b x.
\end{eqnarray}
Changing the sign of $C$ gives
\begin{equation}
\label{1-11.9}
[\hbox{$-$}\hbox{$+$}]_{x-L}^x=\frac{1}{2}(a^2+b^2)+\rme^C b(a-x)+ \rme^{-C} b x.
\end{equation}
In a similar way,
\refstepcounter{equation}
\label{1-11.10}
\addtocounter{equation}{-1}
\numparts
\begin{eqnarray}
\fl
[\hbox{$+$}\hbox{$-$}\hbox{$+$}]_{x-L}^x
=\frac{a}{6}(a^2+3 b^2)+2 b x (x-a) \sinh C+\frac{b}{6}(3 a^2+b^2)\rme^C,
\\
\fl
[\hbox{$-$}\hbox{$+$}\hbox{$-$}]_{x-L}^x
=\frac{a}{6}(a^2+3 b^2)-2 b x (x-a) \sinh C+\frac{b}{6}(3 a^2+b^2)\rme^{-C},
\end{eqnarray}
\endnumparts
\refstepcounter{equation}
\label{1-11.11}
\addtocounter{equation}{-1}
\numparts
\begin{eqnarray}
\fl
[\hbox{$+$}\hbox{$-$}\hbox{$+$}\hbox{$-$}]_{x-L}^x
=&\frac{1}{24}(a^4+6 a^2b^2+b^4)
+\frac{1}{3}[b(3 a^2+b^2)x-6abx^2+4bx^3]\sinh C
\nonumber \\
& \qquad +\frac{ab}{6}(a^2+b^2)\rme^{-C},
\\
\fl
[\hbox{$-$}\hbox{$+$}\hbox{$-$}\hbox{$+$}]_{x-L}^x
=&\frac{1}{24}(a^4+6 a^2b^2+b^4)
-\frac{1}{3}[b(3 a^2+b^2)x-6abx^2+4bx^3]\sinh C
\nonumber \\
& \qquad +\frac{ab}{6}(a^2+b^2)\rme^C.
\end{eqnarray}
\endnumparts
Substituting (\ref{1-11.11}) into (\ref{1-8.16}), we have
\begin{equation}
\label{1-11.12}
Q=\frac{1}{12}(a^4+6 a^2b^2+b^4)+\frac{ab}{3}(a^2+b^2)\cosh C.
\end{equation}
For $0<y\leq x <a$, it is obvious that $[\hbox{$+$}]_y^x=x-y$. 
The low-energy expansion of $G_{\rm S}(x,y;k)$ for $0<y\leq x<a$ is obtained by substituting the above expressions into (\ref{1-10.4}) (and (\ref{1-9.7})). The result of the expansion up to order $k^2$ is plotted in figures~1(b) and (c) (the graphs on the right). It can be seen that this expansion gives a very good approximation in the lowest band except near the band edge.


\section{Conclusion}
The low-energy expansion formula for the reflection coefficient introduced in \cite{analysis} and \cite{low} for non-periodic potentials (equations (\ref{1-3.11}), (\ref{1-3.12}) and (\ref{1-3.13}) of the present paper) is applicable to periodic potentials as well, if only we replace (\ref{1-3.3}) by (\ref{1-4.3}). 
In the periodic case, the general expression for the $n$th-order coefficient of the expansion is obtained as (\ref{1-8.19}). The expansion of the Green function takes the form of (\ref{1-10.1}), with the first four coefficients given by (\ref{1-10.4}).
The method of this paper can be extended to more general potentials which are asymptotically periodic at spatial infinity, and this will be discussed elsewhere.


\appendix


\section{Derivation of (\ref{2-2.10})}
As noted in section~2, the functions $\alpha(x,x';\pm k)+\beta(x,x';\pm k)$ are  solutions of the Schr\"odinger equation (\ref{1-1.3}). If $f$ is identically zero (i.e. $V=\hbox{constant}$), then obviously $\alpha(x,x';\pm k)=\rme^{\mp \rmi k(x-x')}$ and $\beta(x,x';\pm k)=0$. Hence,
\begin{equation}
\label{3-a.1}
\alpha(x,x';\pm k)+\beta(x,x';\pm k)=\rme^{\mp \rmi k(x-x')}
\quad \hbox{if $f=0$}.
\end{equation}
Note that definition (\ref{1-2.3}) is not restricted to $x\geq x'$. 
Since $U(x',x;k)$ is the inverse of the matrix $U(x,x';k)$, we have
\begin{equation}
\label{3-a.2}
\alpha(x',x;\pm k)=\alpha(x,x',\mp k), \qquad 
\beta(x',x;\pm k)=-\beta(x,x';\pm k).
\end{equation}

Let us replace $f(x)$ by $f_{x_1,x_2}(x)$.
If we define
\begin{equation}
\chi_1(x) \equiv \alpha(x,x_1;k)+\beta(x,x_1;k),
\end{equation}
then $\chi_1(x)$ is a solution of the Schr\"odinger equation (\ref{2-2.9}). Since $f_{x_1,x_2}(x)=0$ for $x<x_1$, it is obvious from (\ref{3-a.1}) that
\begin{equation}
\label{3-a.4}
\chi_1(x)=e^{-\rmi k(x-x_1)} \quad  \hbox{for $x<x_1$}. 
\end{equation}
From (\ref{1-2.5}) we have
\begin{eqnarray}
\alpha(x,x_1;k)=\alpha(x,x_2;k)\alpha(x_2,x_1;k)+\beta(x,x_2;-k)\beta(x_2,x_1;k),
\nonumber \\
\beta(x,x_1;k)=\beta(x,x_2;k)\alpha(x_2,x_1;k)+\alpha(x,x_2;-k)\beta(x_2,x_1;k).
\end{eqnarray}
 Using this, we can write
\begin{eqnarray}
\label{3-a.6}
\chi_1(x)&=\alpha(x_2,x_1;k)[\alpha(x,x_2;k)+\beta(x,x_2;k)]
\nonumber \\
& \qquad +\beta(x_2,x_1;k)[\alpha(x,x_2;-k)+\beta(x,x_2;-k)].
\end{eqnarray}
Since $f_{x_1,x_2}(x)=0$ for $x>x_2$, from (\ref{3-a.1}) and (\ref{3-a.6}) we find
\begin{equation}
\label{3-a.7}
\chi_1(x)=\alpha(x_2,x_1;k) \rme^{-\rmi k(x-x_2)}+\beta(x_2,x_1;k) \rme^{\rmi k (x-x_2)} \quad \hbox{for $x>x_2$}. 
\end{equation}
If we divide the right-hand sides of (\ref{3-a.4}) and (\ref{3-a.7}) by $\alpha(x_2,x_1;k)$, they coincide with the right-hand sides of (\ref{2-2.10a}). Hence, we can see that $\psi_1(x)\equiv \chi_1(x)/\alpha(x_2,x_1;k)$ is the solution of the Schr\"odinger equation (\ref{2-2.9}) satisfying (\ref{2-2.10a}). 

In a similar way, we define 
\begin{equation}
\chi_2(x)\equiv \alpha(x,x_2;-k) +\beta(x,x_2;-k).
\end{equation}
By using 
\begin{eqnarray}
\alpha(x,x_2;-k)=\alpha(x,x_1;-k)\alpha(x_2,x_1;k)-\beta(x,x_2;-k)\beta(x_2,x_1;-k), 
\nonumber \\
\beta(x,x_2;-k)=\beta(x,x_1;-k)\alpha(x_2,x_1;k)-\alpha(x,x_2;-k)\beta(x_2,x_1;-k),
\end{eqnarray}
we can rewrite $\chi_2$ as
\begin{eqnarray}
\chi_2(x)&=\alpha(x_2,x_1;k)[\alpha(x,x_1;-k)+\beta(x,x_1;-k)]
\nonumber \\
& \qquad -\beta(x_2,x_1;-k)[\alpha(x,x_1;k)+\beta(x,x_1;k)].
\end{eqnarray}
It is obvious that $\psi_2(x)\equiv \chi_2(x)/\alpha(x_2,x_1;k)$ is the solution of the Schr\"odinger equation satisfying (\ref{2-2.10b}).


\section{Analyticity of $\boldsymbol{\alpha (k)}$, $\boldsymbol{\beta (k)}$, and the reflection coefficients }
We define the functions $\tilde \alpha^\pm$ and $\tilde \beta^\pm$ by
\begin{eqnarray}
\label{2-b.1}
\fl
\left(
\begin{array}{cc}
\tilde \alpha^+(x,x';k) & \tilde \beta^-(x,x';k) \\
\tilde \beta^+(x,x';k) & \tilde \alpha^-(x,x';k) \\
\end{array}
\right)
\equiv
\frac{1}{2}
 \left(
\begin{array}{cc}
1 & 1 \\
-1 & 1 \\
\end{array}
\right)
 \left(
\begin{array}{cc}
\alpha(k) & \beta(-k) \\
\beta(k) & \alpha(-k) \\
\end{array}
\right)
 \left(
\begin{array}{cc}
1 & -1 \\
1 & 1 \\
\end{array}
\right)
\nonumber \\
\fl \quad 
=
\frac{1}{2}
\left(
\begin{array}{cc}
\alpha(k)+\alpha(-k)+\beta(k)+\beta(-k) &  -\alpha(k)+\alpha(-k)-\beta(k)+\beta(-k) \\
-\alpha(k)+\alpha(-k)+\beta(k)-\beta(-k) & \alpha(k)+\alpha(-k)-\beta(k)-\beta(-k) \\
\end{array}
\right),
\end{eqnarray}
where $\alpha(k)$ and $\beta(k)$ stand for $\alpha(x,x';k)$ and $\beta(x,x';k)$.
From (\ref{1-2.1}), it follows that
\begin{eqnarray}
\label{2-b.2}
\fl
\frac{\partial}{\partial x}
\left(
\begin{array}{cc}
\tilde \alpha^+(x,x';k) & \tilde \beta^-(x,x';k) \\
\tilde \beta^+(x,x';k) & \tilde \alpha^-(x,x';k) \\
\end{array}
\right)
=
\left(
\begin{array}{cc}
f(x) & \rmi k \\
\rmi k & -f(x)\\
\end{array}
\right)
\left(
\begin{array}{cc}
\tilde \alpha^+(x,x';k) & \tilde \beta^-(x,x';k) \\
\tilde \beta^+(x,x';k) & \tilde \alpha^-(x,x';k) \\
\end{array}
\right).
\nonumber \\
\end{eqnarray}
Using (\ref{1-1.2}), we can rewrite (\ref{2-b.2}) as
\begin{eqnarray}
\label{2-b.3}
\frac{\partial}{\partial x}
\Bigl[\rme^{V(x)/2}\tilde \alpha^+(x,x';k)\Bigr]
&= \rmi k \rme^{V(x)}
\Bigl[\rme^{-V(x)/2}\tilde \beta^+(x,x';k)\Bigr],
\nonumber \\
\frac{\partial}{\partial x}
\Bigl[\rme^{-V(x)/2}\tilde \beta^+(x,x';k)\Bigr]
&= \rmi k \rme^{-V(x)}
\Bigl[\rme^{V(x)/2}\tilde \alpha^+(x,x';k)\Bigr].
\end{eqnarray}
The initial conditions for these equations are
$\tilde \alpha^+(x',x';k)=1$, $\tilde \beta^+(x',x';k)=0$. 
Equations (\ref{2-b.3}) with these initial conditions are equivalent to the integral equations
\begin{eqnarray}
\rme^{V(x)/2} \tilde \alpha^+(x,x';k)
=\rme^{V(x')/2}
+\rmi k \int_{x'}^x \rme^{V(z)}
\Bigl[\rme^{-V(z)/2}\tilde \beta^+(z,x';k)\Bigr]\,\rmd z,
\nonumber \\
\rme^{-V(x)/2} \tilde \beta^+(x,x';k)
=\rmi k \int_{x'}^x \rme^{-V(z)}
\Bigl[\rme^{V(z)/2}\tilde \alpha^+(z,x';k)\Bigr]\,\rmd z.
\end{eqnarray}
These equations can be solved by iteration.
The solution is formally expressed as
\begin{eqnarray}
\label{2-b.5}
\tilde \alpha^+(x,x';k)=\rme^{-[V(x)-V(x')]/2}
\sum_{m=0}^\infty
(\rmi k)^{2m}I^+_{2m}(x,x'),
\nonumber \\
\tilde \beta^+(x,x';k)=\rme^{[V(x)+V(x')]/2}
\sum_{m=0}^\infty
(\rmi k)^{2m+1}I^+_{2m+1}(x,x'),
\end{eqnarray}
where $I^+_{2m}$ and $I^+_{2m+1}$ are defined recursively by
\begin{equation}
\fl
I^+_{2m}(x,x')=\int_{x'}^x\rme^{V(z)}I^+_{2m-1}(z,x')\,\rmd z,
\qquad
I^+_{2m+1}(x,x')=\int_{x'}^x\rme^{-V(z)}I^+_{2m}(z,x')\,\rmd z,
\end{equation}
with $I^+_0=1$.
It is easy to see that $I^+_1(x,x')=[\hbox{$-$}]_{x'}^x$, $I^+_2(x,x')=[\hbox{$-$$+$}]_{x'}^x$, $I^+_2(x,x')=[\hbox{$-$$+$$-$}]_{x'}^x$, and so on. Namely,
\begin{equation}
\label{2-b.7}
\fl
I^+_{2m}(x,x')=[\hbox{$-$$+$$-$$+$\,$\cdots$\,$-$}$+$]_{x'}^x,
\qquad
I^+_{2m+1}(x,x')=[\hbox{$-$$+$$-$$+$\,$\cdots$\,$+$$-$}]_{x'}^x.
\end{equation}
In the first equation of (\ref{2-b.7}), there are $2m$ alternating plus and minus signs in the expression on the right-hand side. In the second equation, there are $2m+1$ alternating signs.

We assume that both $x$ and $x'$ are finite. As we are assuming that $V(x)$ is finite for any finite $x$, there exists a constant number $C$ such that
\begin{equation}
\rme^{-V(z)}<C , \qquad \rme^{V(z)}<C \qquad \hbox{for} \quad x'\leq z \leq x.
\end{equation}
From definition (\ref{1-2.15}), we can easily see that
\begin{equation}
I^+_{2m}(x,x')<\frac{C^{2m}}{(2m)!}, \qquad I^+_{2m+1}(x,x')<\frac{C^{2m+1}}{(2m+1)!}.
\end{equation}
Therefore,
\begin{eqnarray}
\sum_{m=0}^\infty
\left\vert(\rmi k)^{2m}I_{2m}^+(x,x')\right\vert
<
\sum_{m=0}^\infty \vert k \vert^{2m}\frac{C^{2m}}{(2m)!}
=\cosh \left( C \vert k \vert \right),
\nonumber \\
\sum_{m=0}^\infty
\left\vert(\rmi k)^{2m+1}I_{2m+1}^+(x,x')\right\vert
<
\sum_{m=0}^\infty \vert k \vert^{2m+1}\frac{C^{2m+1}}{(2m+1)!}
=\sinh \left( C \vert k \vert \right).
\end{eqnarray}
Thus, the infinite series on the right-hand sides of (\ref{2-b.5}) are absolutely convergent for any $k$. This means that $\tilde \alpha^+(x,x';k)$ and $\tilde \beta^+(x,x;k)$ are entire functions of $k$. 

In the same way, we have
\begin{eqnarray}
\label{2-b.11}
\tilde \alpha^-(x,x';k)=\rme^{[V(x)-V(x')]/2}
\sum_{m=0}^\infty
(\rmi k)^{2m}I^-_{2m}(x,x'),
\nonumber \\
\tilde \beta^-(x,x';k)=\rme^{-[V(x)+V(x')]/2}
\sum_{m=0}^\infty
(\rmi k)^{2m+1}I^-_{2m+1}(x,x'),
\end{eqnarray}
where
\begin{equation}
\label{2-b.12}
\fl
I^-_{2m}(x,x')=[\hbox{$+$$-$$+$$-$\,$\cdots$\,$+$$-$}]_{x'}^x,
\qquad
I^-_{2m+1}(x,x')=[\hbox{$+$$-$$+$$-$\,$\cdots$\,$-$$+$}]_{x'}^x.
\end{equation}
Equations (\ref{2-b.11}) and (\ref{2-b.12}) are obtained from (\ref{2-b.5}) and (\ref{2-b.7}) by changing the sign of the potential $V$.
Just like $\tilde \alpha^+$ and $\tilde \beta^+$, the functions $\tilde \alpha^-$ and $\tilde \beta^-$ are entire functions of $k$.
Inverting (\ref{2-b.1}), we can express $\alpha(\pm k)$ and $\beta(\pm k)$ in terms of $\tilde \alpha^\pm$ and $\tilde \beta^\pm$ as
\begin{eqnarray}
\label{2-b.13}
\fl
\left(
\begin{array}{cc}
\alpha(x,x';k) & \beta(x,x';-k) \\
\beta(x,x';k) & \alpha(x,x';-k) \\
\end{array}
\right)
=
\frac{1}{2}
\left(
\begin{array}{cc}
\tilde \alpha^+ + \tilde \alpha^- -\tilde \beta^+ -\tilde \beta^- & 
\tilde \alpha^+ - \tilde \alpha^- -\tilde \beta^+ +\tilde \beta^- \\
\tilde \alpha^+ - \tilde \alpha^- +\tilde \beta^+ -\tilde \beta^- & 
\tilde \alpha^+ + \tilde \alpha^- +\tilde \beta^+ +\tilde \beta^-\\
\end{array}
\right).
\nonumber \\
\end{eqnarray}
Obviously $\alpha(x,x';k)$ and $\beta(x,x';k)$ are entire functions of $k$ as long as both $x$ and $x'$ are finite.
We obtain (\ref{1-5.10}) by substituting (\ref{2-b.5}) and (\ref{2-b.11}) into (\ref{2-b.13}).

It is shown in appendix~D that the reflection coefficient $R_r(x,-\infty;k)$ $({\rm Im}\,k>0$)  for periodic potentials can be expressed in terms of $\alpha(k)\equiv \alpha(x,x-L;k)$ and $\beta(k)\equiv \beta(x,x-L;k)$ as
\begin{equation}
\label{2-b.14}
R_r(x,-\infty;k) = \frac{-\beta(k)}{[1/\lambda(k)]-\alpha(k)}
\end{equation}
with $\lambda(k)=Y(k) -\rmi \sqrt{1-[Y(k)]^2}$ and $Y(k)=[\alpha(k)+\alpha(-k)]/2$.
Since $\alpha(k)$ and $\beta(k)$ are entire functions, $R_r(x,-\infty;k)$ may possibly have singularities only where $1/\lambda(k)=\alpha(k)$ or $Y(k)=\pm 1$. 

It can be shown that $\vert \tau(x_2,x_1;k) \vert< 1$ for ${\rm Im}\,k>0$ and $x_1<x_2$ (see section XIII of \cite{theory}).
This implies that $\vert \alpha(k) \vert>1$ for ${\rm Im}\,k>0$. 
Since $\vert \lambda(k)\vert >1$ for ${\rm Im}\,k>0$, 
the equality $1/\lambda(k)=\alpha(k)$ can never hold in ${\rm Im}\,k>0$.
If $Y(k)=\pm 1$, then $\vert \lambda(k) \vert =1$. 
This cannot hold for ${\rm Im}\,k>0$, either. 
(If $\vert \lambda(k) \vert =1$, there exist two independent solutions of the Schr\"odinger equation which remain finite as $x \to \pm \infty$. This can happen for ${\rm Im}\,k=0$, but not for ${\rm Im}\, k>0$.)  
Thus, neither $1/\lambda(k)=\alpha(k)$ nor $Y(k)=\pm 1$ can hold in ${\rm Im}\,k>0$, and so $R_r(x,-\infty;k)$ is analytic in the upper half plane (${\rm Im}\,k>0$). 
Note that $R_r(x,-\infty;k)$ has branch point singularities on the real axis. These branch points correspond to the edges of the bands. 

The power series expansion of $R_r(x,-\infty;k)$ in terms of $k$ is convergent for sufficiently small $\vert k \vert$ if the right-hand side of (\ref{2-b.14}) is analytic in some neighborhood of the origin. Since
\begin{eqnarray}
[1/\lambda(k)]-\alpha(k)=2 \rmi k L_0 \cosh^2\frac{V(x)-V_0}{2}+O(k^3),
\nonumber \\
1-[Y(k)]^2=k^2L_0^2+O(k^4),
\end{eqnarray}
there exists a number $C$ such that $[1/\lambda(k)]-\alpha(k)\neq 0$ and $1-\vert Y(k)\vert^2 \neq 0$ for $0<\vert k \vert <C$.
Although $[1/\lambda(k)]-\alpha(k)=0$ at $k=0$, this zero is canceled by the zero of $\beta(k)$ in the numerator (see equation (\ref{1-5.15})), and so the right-hand side of (\ref{2-b.14}) does not have a pole at $k=0$. 
The origin is not a branch point, either, since $\sqrt{1-\vert Y(k)\vert^2} \simeq k L_0$  for $k \simeq 0$. Therefore $R_r(x,-\infty;k)$ is analytic, and the power series expansion in terms of $k$ is convergent, in $\vert k \vert <C$. 

Since $\bar R_r(x,-\infty;W;0)=-\tanh \frac{W-V_0}{2}\neq \infty$ (see (\ref{1-8.2})), we know that $\bar R_r$ does not have a pole at the origin. As $R_r(x,-\infty;k)$ is analytic in a neighborhood of $k=0$, it is obvious from definition (\ref{1-2.12b}) that $\bar R_r(x,-\infty;W;k)$, too, is analytic in some neighborhood of the origin.
Thus, the expansion of $\bar R_r(x,-\infty;W;k)$ in terms of $k$ is convergent for sufficiently small $\vert k \vert$.


\section{Relation with the Weyl-Titchmarsh $\boldsymbol{m}$-function}
A natural way to define the transmission and reflection coefficients (for finite intervals)  for the Schr\"odinger equation is to consider the truncated potential
\begin{equation}
V^{\rm S}_{x_1,x_2}(x)\equiv
\cases{
0 & $(x<x_1)$ \\
V_{\rm S}(x) & $(x_1\leq x \leq x_2)$ \\
0 & $(x_2<x)$.
}
\end{equation}
This $V^{\rm S}_{x_1,x_2}$ is different from $f^2_{x_1,x_2}+f'_{x_1,x_2}$ (which appears in (\ref{2-2.9})), since the latter contains delta functions at $x=x_1$ and $x=x_2$ (unless $f(x_1)=0$ or $f(x_2)=0$). The Schr\"odinger equation
\begin{equation}
-\frac{\rmd^2 \psi}{\rmd x^2}
+V^{\rm S}_{x_1,x_2}\psi=k^2 \psi
\end{equation}
has two independent solutions of the forms
\begin{equation}
\label{2-c.3}
\psi_1(x)=
\cases{
\tau^{\rm S}(x_2,x_1;k)\, \rme^{-\rmi k(x-x_1)}  &$(x<x_1)$\\
\rme^{-\rmi k(x-x_2)} + R_r^{\rm S}(x_2,x_1;k)\, \rme^{\rmi k(x-x_2)} &$(x>x_2)$,\\
}
\end{equation}
\begin{equation}
\psi_2(x)=
\cases{
\rme^{\rmi k(x-x_1)} + R_l^{\rm S}(x_2,x_1;k)\, \rme^{-\rmi k(x-x_1)} &$(x<x_1)$\\
\tau^{\rm S}(x_2,x_1;k)\, \rme^{\rmi k(x-x_2)} &  $(x>x_2)$. \\
}
\end{equation}
This defines the transmission and reflection coefficients $\tau^{\rm S}$, $R_r^{\rm S}$ and $R_l^{\rm S}$ for the Schr\"odinger equation.
Here we assume that ${\rm Im}\,k>0$.
When $V_{\rm S}$ belongs to the so-called limit point case \cite{coddington}, we can let $x_1 \to -\infty$ or $x_2 \to +\infty$ to obtain the reflection coefficients for semi-infinite intervals. 
As shown in \cite{low}, the reflection coefficients $R_r^{\rm S}(x,-\infty;k)$ and $R_l^{\rm S}(\infty,x;k)$ are related to $R_r(x,-\infty;k)$ and $R_l(\infty,x;k)$ by
\begin{eqnarray}
\label{2-c.5}
\frac{R_r(x,-\infty;k)}{1+R_r(x,-\infty;k)}
=\frac{R_r^{\rm S}(x,-\infty;k)}{1+R_r^{\rm S}(x,-\infty;k)}
-\frac{f(x)}{2 \rmi k},
\nonumber \\
\label{2-c.6}
\frac{R_l(\infty,x;k)}{1+R_l(\infty,x;k)}
=\frac{R_l^{\rm S}(\infty,x;k)}{1+R_l^{\rm S}(\infty,x;k)}
+\frac{f(x)}{2 \rmi k}.
\end{eqnarray}

Let $\eta_1$ and $\eta_2$ be two independent solutions of the Schr\"odinger equation (\ref{1-1.3}) satisfying the conditions
\begin{equation}
\eta_1(x_0)=1,  \qquad \eta'_1(x_0)=0, \qquad  \eta_2(x_0)=0,  \qquad \eta'_2(x_0)=1,
\end{equation}
with some fixed $x_0$. And let $\chi_\pm$ denote the solutions of (\ref{1-1.3}) satisfying $\chi_+(\infty)=0$ and $\chi_-(-\infty)=0$ (and which are are square integrable near $+ \infty$ and $-\infty$, respectively). These functions can be expressed as linear combinations of $\eta_1$ and $\eta_2$. We write
\begin{equation}
\chi_\pm(x) = \eta_1(x) \pm m_\pm\, \eta_2(x).
\end{equation}
This is the definition of the Weyl-Titchmarsh functions $m_+(x_0,k)$ and $m_-(x_0,k)$, which are regarded as functions of $x_0$ and $k$. 

Let us suppose that $V_{\rm S}(x)=0$ for $x>x_0$. Then, 
\begin{equation}
\fl
\eta_1(x)=\cos [k (x-x_0)], \qquad \eta_2(x)= \frac{1}{k} \sin [k (x-x_0)] 
\qquad \hbox{for $x>x_0$},
\end{equation}
and so
\begin{eqnarray}
\fl
\chi_-(x)&=\cos [k (x-x_0)] - \frac{m_-(x_0,k)}{k} \sin [k (x-x_0)]
\nonumber \\
\fl
&=\frac{1}{2k}[k+ \rmi m_-(x_0,k)]\,\rme^{\rmi k (x-x_0)}+\frac{1}{2k}[k- \rmi m_-(x_0,k))]\,\rme^{-\rmi k (x-x_0)} 
\quad \hbox{for $x>x_0$}.
\end{eqnarray}
Comparing this with (\ref{2-c.3}), we find \cite{pavlov, rybkin}
\begin{equation}
\label{2-c.10}
\fl
R_r^{\rm S}(x_0,-\infty;k)=\frac{k+\rmi m_-(x_0,k)}{k-\rmi m_-(x_0,k)}, \qquad
m_-(x_0,k)=\rmi k - 2\rmi k\frac{R_r^{\rm S}(x_0,-\infty;k)}{1+R_r^{\rm S}(x_0,-\infty;k)}.
\end{equation}
In the same way,
\begin{equation}
\label{2-c.11}
\fl
R_l^{\rm S}(\infty,x_0;k)=\frac{k+\rmi m_+(x_0,k)}{k-\rmi m_+(x_0,k)}, \qquad
m_+(x_0,k)=\rmi k - 2\rmi k\frac{R_l^{\rm S}(\infty,x_0;k)}{1+R_l^{\rm S}(\infty,x_0;k)}.
\end{equation}
From (\ref{2-c.5}), (\ref{2-c.10}) and (\ref{2-c.11}), we obtain
\begin{equation}
\fl
S_r(x,k)=\frac{\rmi}{2k} [m_-(x,k)+f(k)]+\frac{1}{2}, \qquad
S_l(x,k)=\frac{\rmi}{2k} [m_+(x,k)-f(k)]+\frac{1}{2},
\end{equation}
and
\begin{equation}
S(x,k)=\frac{\rmi}{2k}[m_+(x,k)+m_-(x,k)]+1.
\end{equation}


\section{Exact expressions of $\boldsymbol{R}_r$, $\boldsymbol{R}_l$ and $\boldsymbol{S}$}
The function $\alpha(x_2,x_1;k)$ satisfies the differential equations
\begin{eqnarray}
\label{1-a.1}
\frac{\partial}{\partial x_2}\alpha(x_2,x_1;k)=-\rmi k\alpha(x_2,x_1;k)+f(x_2)\beta(x_2,x_1;k),
\nonumber \\
\frac{\partial}{\partial x_1}\alpha(x_2,x_1;k)=\rmi k\alpha(x_2,x_1;k)-f(x_1)\beta(x_2,x_1;-k).
\end{eqnarray}
(The first equation of (\ref{1-a.1}) is a component of (\ref{1-2.1}). 
The second equation is obtained by using (\ref{3-a.2}) of appendix~A.) 
Hence, if $f$ is periodic,
\begin{equation}
\label{1-a.2}
\frac{\partial}{\partial x}\alpha(x,x-L;k)=f(x)\left[\beta(x,x-L;k)-\beta(x,x-L;-k)\right],
\end{equation}
and so
\begin{equation}
\label{1-a.3}
\frac{\partial}{\partial x}[\alpha(x,x-L;k)+\alpha(x,x-L;-k)]=0.
\end{equation}
Thus, $Y(k)$ defined by (\ref{1-5.7}) is independent of $x$. We define
\begin{equation}
\label{1-a.4}
Z(k) \equiv \left\{1-[Y(k)]^2\right\}^{1/2}
\end{equation}
to write (\ref{1-5.6}) as 
\begin{equation}
\label{1-a.5}
\lambda(k)=Y(k)-\rmi Z(k), \qquad 1/\lambda(k)=Y(k)+\rmi Z(k).
\end{equation}
The branch of the square root in (\ref{1-a.4}) is chosen such that $\vert \lambda(k) \vert>1$ for ${\rm Im}\,k>0$. With this choice of the square root,  $Z(k)$ is an odd function of $k$ whereas $Y(k)$ is an even function (see (\ref{1-5.12})).
So, from (\ref{1-a.5}) we find $1/\lambda(k)=\lambda(-k)$. 

Hereafter we write $\alpha(k)$ and $\beta(k)$ in place of $\alpha(x,x-L;k)$ and $\beta(x,x-L;k)$, respectively. Substituting the elements of (\ref{1-5.8}) into (\ref{1-2.7b}) gives
\begin{eqnarray}
R_r(x,x-n L;k)=
\frac{-\beta(k)(1-\gamma^n)}
{[1/\lambda(k)]-\alpha(k)-\gamma^n[\lambda(k)-\alpha(k)]},
\nonumber \\
\fl
R_l(x+n L,x;k)=
R_l(x,x-n L;k)=
\frac{\beta(-k)(1-\gamma^n)}
{[1/\lambda(k)]-\alpha(k)-\gamma^n[\lambda(k)-\alpha(k)]},
\end{eqnarray}
where $\gamma\equiv [\lambda(k)]^{-2}$.
Since $\vert \gamma \vert <1$ for ${\rm Im}\,k>0$, we can take the limit $n\to \infty$ of these expressions and obtain
\begin{eqnarray}
\label{1-a.6}
R_r(x,-\infty;k)=\frac{-\beta(k)}{[1/\lambda(k)]-\alpha(k)}
=\frac{2 \beta(k)}{\alpha(k)-\alpha(-k)-2 \rmi Z(k)},
\nonumber \\
R_l(\infty,x;k)=\frac{\beta(-k)}{[1/\lambda(k)]-\alpha(k)}
=\frac{-2 \beta(-k)}{\alpha(k)-\alpha(-k)-2 \rmi Z(k)}.
\end{eqnarray}
Substituting into (\ref{1-2.8}), and also using (\ref{1-2.4}), we have
\begin{eqnarray}
\label{1-a.7}
\fl
S_r(x,k)=\frac{2\beta(k)}{\alpha(k)-\alpha(-k)+2\beta(k)-2\rmi Z(k)}
=\frac{1}{2}
\frac{\alpha(k)-\alpha(-k)+2\beta(k)+2\rmi Z(k)}{\alpha(k)-\alpha(-k)+\beta(k)-\beta(-k)},
\nonumber \\
\fl
S_l(x,k)=\frac{-2\beta(-k)}{\alpha(k)-\alpha(-k)-2\beta(-k)-2\rmi Z(k)}
=\frac{1}{2}
\frac{\alpha(k)-\alpha(-k)-2\beta(-k)+2\rmi Z(k)}{\alpha(k)-\alpha(-k)+\beta(k)-\beta(-k)}.
\nonumber \\
\fl
\end{eqnarray}
Since $Z(-k)=-Z(k)$, it is obvious that $S_l(x,k)=S_r(x,-k)$.
The expression of $S$ is obtained from (\ref{1-a.7}) as
\begin{equation}
\label{1-a.8}
S(x,k)=1+\frac{2 \rmi Z(k)}{\alpha(k)-\alpha(-k)+\beta(k)-\beta(-k)}.
\end{equation}
By definition,
the above expressions hold for ${\rm Im}\,k=0$ as well.

When $k$ is a real number, $[\alpha(k)]^* =\alpha(-k)$, $[\beta(k)]^*=\beta(-k)$, and $Y(k)={\rm Re}\,\alpha(k)$. 
If $[Y(k)]^2>1$, the two eigenvalues of $U(x,x-L;k)$ are real, and $\vert \lambda(k) \vert>1$. If $[Y(k)]^2<1$, then $\vert \lambda(k) \vert =1$. 
This means that the energy lies is in a band if $[Y(k)]^2<1$, and in a gap if $[Y(k)]^2>1$ (see, for example, \cite{coddington}).


\section{Details of  the calculation leading to (\ref{1-5.17}) and (\ref{1-5.18})}
To simplify the expressions, we write
\begin{eqnarray}
\label{2-e.1}
\fl
\delta \equiv \frac{V(z)-V(x)}{2}, \quad \  
\omega \equiv \frac{V(z)+V(x)}{2}, \quad \  
\theta \equiv \frac{V(x)-V_0}{2}, \quad \  
\varphi \equiv  \frac{W-V(x)}{2},
\end{eqnarray}
and
\begin{equation}
\label{2-e.2}
b \equiv \frac{1}{2} \left(\,[\hbox{$-$}\hbox{$+$}]_{x-L}^{x}-\,[\hbox{$+$}\hbox{$-$}]_{x-L}^{x}\right).
\end{equation}
Then, (\ref{2-5.16}) and (\ref{2-5.17}) read
\begin{eqnarray}
\label{2-e.3}
\fl
\alpha(x,x-nL;\pm k)&=(\cosh^2 \theta)\, \lambda^{\pm n}
-(\sinh^2 \theta) \,\lambda^{\mp n} + O(k^2),
\nonumber \\
\fl
\beta(x,x-nL;\pm k)&=
\left(
\mp \sinh \theta \cosh \theta -\rmi k \frac{b}{2L_0} 
\right)
\left(
\lambda^n-\lambda^{-n}
\right)
+O(k^2),
\\
\label{2-e.4}
\fl
\alpha(x-nL,z;\pm k)&=\cosh \delta \mp \frac{\rmi k}{2}
\left(\rme^\omega\,[\hbox{$-$}]_{z}^{x-nL}+\rme^{-\omega}\,[\hbox{$+$}]_{z}^{x-nL}\right)
+O(k^2),
\nonumber \\
\fl
\beta(x-nL,z;\pm k)&=\sinh \delta \pm \frac{\rmi k}{2}
\left(\rme^\omega\,[\hbox{$-$}]_{z}^{x-nL}-\rme^{-\omega}\,[\hbox{$+$}]_{z}^{x-nL}\right)
+O(k^2).
\end{eqnarray}
Substituting (\ref{2-e.3}) and (\ref{2-e.4}) into (\ref{2-5.18}) yields
\begin{eqnarray}
\label{2-e.5}
\alpha(x,z;\pm k)
&=\cosh \theta \cosh (\theta +\delta) \, \lambda^{\pm n}
 -\sinh \theta \sinh (\theta +\delta) \, \lambda^{\mp n}
\nonumber \\
& \qquad
\mp \frac{\rmi k}{2} \cosh \theta 
\left(\rme^{\omega-\theta}\,[\hbox{$-$}]_{z}^{x-nL}+\rme^{\theta-\omega}\,[\hbox{$+$}]_{z}^{x-nL}\right) \,\lambda^{\pm n}
\nonumber \\
& \qquad
\mp \frac{\rmi k}{2} \sinh \theta 
\left(\rme^{\omega-\theta}\,[\hbox{$-$}]_{z}^{x-nL}-\rme^{\theta-\omega}\,[\hbox{$+$}]_{z}^{x-nL}\right) \,\lambda^{\mp n}
\nonumber \\
& \qquad
-\rmi k \frac{b}{2L_0} \sinh \delta  \left(\lambda^n-\lambda^{-n}\right)+O(k^2),
\nonumber
\end{eqnarray}
\begin{eqnarray}
\beta(x,z;\pm k)
&=-\sinh \theta \cosh (\theta +\delta) \, \lambda^{\pm n}
 +\cosh \theta \sinh (\theta +\delta) \, \lambda^{\mp n}
\nonumber \\
& \qquad
\pm \frac{\rmi k}{2} \sinh \theta 
\left(\rme^{\omega-\theta}\,[\hbox{$-$}]_{z}^{x-nL}+\rme^{\theta-\omega}\,[\hbox{$+$}]_{z}^{x-nL}\right) \,\lambda^{\pm n}
\nonumber \\
& \qquad
\pm \frac{\rmi k}{2} \cosh \theta 
\left(\rme^{\omega-\theta}\,[\hbox{$-$}]_{z}^{x-nL}-\rme^{\theta-\omega}\,[\hbox{$+$}]_{z}^{x-nL}\right) \,\lambda^{\mp n}
\nonumber \\
& \qquad
-\rmi k \frac{b}{2L_0} \cosh \delta  \left(\lambda^n-\lambda^{-n}\right)+O(k^2).
\end{eqnarray}
Equations (\ref{2-e.5}) can be written in the form
\begin{eqnarray}
\label{2-e.6}
\fl
\left(
\begin{array}{cc}
\alpha(x,z;k) & \beta(x,z;-k) \\
\beta(x,z;k) & \alpha(x,z;-k) \\
\end{array}
\right)
&=
\left(
\begin{array}{cc}
\cosh \theta & -\sinh \theta \\
-\sinh \theta & \cosh \theta \\
\end{array}
\right)
\left(
\begin{array}{cc}
A(k) & B(-k) \\
B(k) & A(-k) \\
\end{array}
\right)
\nonumber \\
& \qquad
- C(k) \left(
\begin{array}{cc}
\sinh \delta & \cosh \delta \\
\cosh \delta & \sinh \delta \\
\end{array}
\right),
\end{eqnarray}
where
\begin{eqnarray}
A(\pm k)\equiv \cosh (\theta +\delta) \lambda^{\pm n}
\mp \frac{\rmi k}{2}
\left(\rme^{\omega-\theta}\,[\hbox{$-$}]_{z}^{x-nL}
+\rme^{\theta-\omega}\,[\hbox{$+$}]_{z}^{x-nL}\right)
 \,\lambda^{\pm n},
\nonumber \\
B(\pm k)\equiv \sinh (\theta +\delta) \lambda^{\mp n}
\pm \frac{\rmi k}{2}
\left(\rme^{\omega-\theta}\,[\hbox{$-$}]_{z}^{x-nL}
-\rme^{\theta-\omega}\,[\hbox{$+$}]_{z}^{x-nL}\right)
 \,\lambda^{\mp n},
\nonumber \\
C(k) \equiv \rmi k \frac{b}{2L_0}\left(\lambda^n-\lambda^{-n}\right).
\end{eqnarray}
Therefore,
\begin{eqnarray}
\label{2-e.8}
\fl
\left(
\begin{array}{cc}
\bar \alpha(x,z;W;k) & \bar \beta(x,z;W;-k) \\
\bar \beta(x,z;W;k) & \bar \alpha(x,z;W;-k) \\
\end{array}
\right)
\nonumber \\
\fl
\qquad \qquad \qquad 
=
\left(
\begin{array}{cc}
\cosh \varphi & -\sinh \varphi \\
-\sinh \varphi & \cosh \varphi \\
\end{array}
\right)
\left(
\begin{array}{cc}
\cosh \theta & -\sinh \theta \\
-\sinh \theta & \cosh \theta \\
\end{array}
\right)
\left(
\begin{array}{cc}
A(k) & B(-k) \\
B(k) & A(-k) \\
\end{array}
\right)
\nonumber \\
\fl \qquad \qquad \qquad \qquad
- C(k) 
\left(
\begin{array}{cc}
\cosh \varphi & -\sinh \varphi \\
-\sinh \varphi & \cosh \varphi \\
\end{array}
\right)
\left(
\begin{array}{cc}
\sinh \delta & \cosh \delta \\
\cosh \delta & \sinh \delta \\
\end{array}
\right)
\nonumber \\
\fl \qquad \qquad \qquad
=
\left(
\begin{array}{cc}
\cosh (\theta+\varphi) & -\sinh (\theta+\varphi) \\
-\sinh (\theta+\varphi) & \cosh (\theta+\varphi) \\
\end{array}
\right)
\left(
\begin{array}{cc}
A(k) & B(-k) \\
B(k) & A(-k) \\
\end{array}
\right)
\nonumber \\
\fl \qquad \qquad \qquad \qquad
- C(k) \left(
\begin{array}{cc}
\sinh (\delta -\varphi) & \cosh (\delta -\varphi) \\
\cosh (\delta -\varphi) & \sinh (\delta -\varphi) \\
\end{array}
\right).
\end{eqnarray}
Comparing (\ref{2-e.6}) with (\ref{2-e.8}), we can see that $\bar \alpha$ and $\bar \beta$ are obtained from $\alpha$ and $\beta$, respectively, by the replacements
\begin{equation}
\label{2-e.9}
\theta \to \theta + \varphi, \qquad
\delta \to \delta - \varphi, \qquad
\omega \to \omega + \varphi.
\end{equation}
Namely, 
\begin{eqnarray}
\label{2-e.10}
\fl
\bar \alpha(x,z;W;\pm k)
&=\cosh (\theta+\varphi) \cosh (\theta +\delta) \, \lambda^{\pm n}
 -\sinh (\theta+\varphi) \sinh (\theta +\delta) \, \lambda^{\mp n}
\nonumber \\
\fl
& \qquad
\mp \frac{\rmi k}{2} \cosh (\theta+\varphi) 
\left(\rme^{\omega-\theta}\,[\hbox{$-$}]_{z}^{x-nL}+\rme^{\theta-\omega}\,[\hbox{$+$}]_{z}^{x-nL}\right) \,\lambda^{\pm n}
\nonumber \\
\fl
& \qquad
\mp \frac{\rmi k}{2} \sinh (\theta+\varphi)
\left(\rme^{\omega-\theta}\,[\hbox{$-$}]_{z}^{x-nL}-\rme^{\theta-\omega}\,[\hbox{$+$}]_{z}^{x-nL}\right) \,\lambda^{\mp n}
\nonumber \\
& \qquad
-\rmi k\frac{b}{2L_0} \sinh (\delta-\varphi)  \left(\lambda^n-\lambda^{-n}\right)+O(k^2),
\nonumber 
\end{eqnarray}
\begin{eqnarray}
\fl
\bar \beta(x,z;W;\pm k)
&=-\sinh (\theta+\varphi) \cosh (\theta +\delta) \, \lambda^{\pm n}
 +\cosh (\theta+\varphi) \sinh (\theta +\delta) \, \lambda^{\mp n}
\nonumber \\
\fl
& \qquad
\pm \frac{\rmi k}{2} \sinh (\theta+\varphi)
\left(\rme^{\omega-\theta}\,[\hbox{$-$}]_{z}^{x-nL}+\rme^{\theta-\omega}\,[\hbox{$+$}]_{z}^{x-nL}\right) \,\lambda^{\pm n}
\nonumber \\
\fl
& \qquad
\pm \frac{\rmi k}{2} \cosh (\theta+\varphi)
\left(\rme^{\omega-\theta}\,[\hbox{$-$}]_{z}^{x-nL}-\rme^{\theta-\omega}\,[\hbox{$+$}]_{z}^{x-nL}\right) \,\lambda^{\mp n}
\nonumber \\
\fl
& \qquad
-\rmi k\frac{b}{2L_0} \cosh (\delta-\varphi)  \left(\lambda^n-\lambda^{-n}\right)+O(k^2).
\end{eqnarray}
Note that the replacements (\ref{2-e.9}) are equivalent to $V(x) \to W$. 
This is an expected result if we consider the meaning of the variable $W$ (see \cite{algebraic}). 
Equations (\ref{1-5.17}) are obtained by substituting (\ref{2-e.1}) and (\ref{2-e.2}) into (\ref{2-e.10}).

From (\ref{2-e.10}), we obtain
\begin{eqnarray}
\label{2-e.11}
\fl
\bar \alpha(x,z;W;k)+\bar \beta(x,z;W;-k)
&=\rme^{\theta+\delta} 
\left(
\cosh (\theta +\varphi) \, \lambda^n
 -\sinh (\theta +\varphi) \, \lambda^{- n}
 \right)
\nonumber \\
& \quad
- \rmi k\, \rme^{\omega-\theta} 
\left(
\cosh (\theta+\varphi) \lambda^n + \sinh (\theta+\varphi) \lambda^{-n}
\right)[\hbox{$-$}]_{z}^{x-nL}
\nonumber \\
& \quad
-\rmi k\frac{b}{2 L_0} \,\rme^{\delta-\varphi}  \left(\lambda^n-\lambda^{-n}\right)+O(k^2),
\nonumber 
\end{eqnarray}
\begin{eqnarray}
\fl
\bar \alpha(x,z;W;k)-\bar \beta(x,z;W;-
k)
&=\rme^{-(\theta+\delta)} 
\left(
\cosh (\theta +\varphi) \, \lambda^n
 +\sinh (\theta +\varphi) \, \lambda^{- n}
 \right)
\nonumber \\
& \quad
- \rmi k\, \rme^{\theta-\omega}
\left(
\cosh (\theta+\varphi) \lambda^n - \sinh (\theta+\varphi) \lambda^{-n}
\right)[\hbox{$+$}]_{z}^{x-nL}
\nonumber \\
& \quad
+\rmi k\frac{b}{2L_0} \,\rme^{\varphi-\delta}  \left(\lambda^n-\lambda^{-n}\right)+O(k^2).
\end{eqnarray}
Substituting (\ref{2-e.1}), and using definitions (\ref{1-5.19}), we can write (\ref{2-e.11}) as
\begin{eqnarray}
\label{2-e.12}
\fl
\bar \alpha(x,z;W;k)+\bar \beta(x,z;W;-k)
&= \rme^{[V(z)-V_0]/2}\cosh \case{W-V_0}{2}\, (1+c_0 \gamma^n)\, \lambda^n
\nonumber \\
& \qquad \times
\left(
1-\rmi k 
\frac{1-c_0 \gamma^n}{1+c_0 \gamma^n}
\rme^{V_0}\,[\hbox{$-$}]_z^{x-nL}+ k b_1
\right) +O(k^2),
\nonumber \\
\fl
\bar \alpha(x,z;W;k)-\bar \beta(x,z;W;-k)
&= \rme^{-[V(z)-V_0]/2} \cosh \case{W-V_0}{2}\, (1-c_0 \gamma^n)\, \lambda^n
\nonumber \\
& \qquad \times
\left(
1-\rmi k 
\frac{1+c_0 \gamma^n}{1-c_0 \gamma^n}
\rme^{-V_0}\,[\hbox{$+$}]_z^{x-nL}+ k b_2
\right) +O(k^2),
\nonumber \\
\end{eqnarray}
where the quantities $b_1$ and $b_2$ (which come from the terms involving $b$) are independent of $z$. 
Equations (\ref{1-5.18}) are easily obtained from (\ref{2-e.12}).


\section{Verification of $\mathcal{A}\mathcal{A}^{-1} \boldsymbol{g = g}$ and $\mathcal{A}^{-1}\mathcal{A} \boldsymbol{h = h}$}
Let us verify $\mathcal{A}\mathcal{A}^{-1}g=g$ and $\mathcal{A}^{-1}\mathcal{A}h=h$ using expression (\ref{1-4.3}), or equivalently (\ref{1-5.30}).
We can see that
\begin{eqnarray}
\label{1-b.1}
\fl
\frac{\partial}{\partial x}\int_{x-L}^x  
\left(
\rme^{-W} [\hbox{$+$}]_z^x-\rme^W [\hbox{$-$}]_z^x
\right)
g(z,W)\,\rmd z
\nonumber \\
\fl
\quad
=-\left(
\rme^{-W} [\hbox{$+$}]_{x-L}^x-\rme^W [\hbox{$-$}]_{x-L}^x
\right)g(x-L,W)
 +\left(\rme^{V(x)-W}-\rme^{-V(x)+W}\right)\int_{x-L}^x g(z,W)\,\rmd z
\nonumber \\
\fl
\quad
=2L_0 \sinh(W-V_0) g(x,W),
\end{eqnarray} 
where we have used (\ref{1-2.20}), (\ref{1-5.1}) and (\ref{1-5.2}).
Since the last line of (\ref{1-b.1}) vanishes for $W=V_0$, 
we can insert the operator $\mathcal{D}$ in front of the integral in the first line without changing the result.
Namely,
\begin{equation}
\fl
\mathcal{A}\,\mathcal{D}
\int_{x-L}^x  
\left(
\rme^{-W} [\hbox{$+$}]_z^x-\rme^W [\hbox{$-$}]_z^x
\right)
g(z,W)\,\rmd z
=2L_0 \sinh(W-V_0) g(x,W).
\end{equation}
Dividing both sides by $2L_0 \sinh(W-V_0)$, we obtain $\mathcal{A}\mathcal{A}^{-1}g=g$.

In a similar way we have, by integrating by parts,
\begin{eqnarray}
\label{1-b.2}
\fl
\int_{x-L}^x  
\left(
\rme^{-W} [\hbox{$+$}]_z^x-\rme^W [\hbox{$-$}]_z^x
\right)
\frac{\partial}{\partial z}h(z,W)\,\rmd z
\nonumber \\
=2L_0 \sinh(W-V_0) h(x,W)
-2 \int_{x-L}^x \sinh[W-V(z)]h(z,W)\,\rmd z.
\end{eqnarray} 
We apply the operator $\mathcal{D}$ to both sides of (\ref{1-b.2}).
The first term on the right-hand side does not change. 
For the second term, we can use (\ref{1-2.24}) to write
\begin{eqnarray}
\label{1-b.3}
\fl
\mathcal{D}\int_{x-L}^x \sinh[W-V(z)]h(z,W)\,\rmd z
&=\int_{V_0}^W \rmd W\int_{x-L}^x \rmd z\,
\frac{\partial}{\partial W}\sinh[W-V(z)]h(z,W)
\nonumber \\
&=\int_{V_0}^W \rmd W\int_{x-L}^x \rmd z\,\mathcal{B} h(z,W).
\end{eqnarray}
This vanishes on account of (\ref{1-6.3}). 
Therefore,
\begin{equation}
\fl
\mathcal{D}
\int_{x-L}^x  
\left(
\rme^{-W} [\hbox{$+$}]_z^x-\rme^W [\hbox{$-$}]_z^x
\right)
\mathcal{A}h(z,W)\,\rmd z
=2L_0 \sinh(W-V_0) h(x,W).
\end{equation}
Dividing by $2L_0 \sinh(W-V_0)$ gives $\mathcal{A}^{-1}\mathcal{A}h=h$.


\section{Verification of (\ref{1-6.4})} 
We can write the left-hand side of (\ref{1-6.4}) as
\begin{eqnarray}
\label{1-b.4}
\fl
\int_{x-L}^x \mathcal{B}\mathcal{A}^{-1}g(z,W)\,\rmd z
=\frac{1}{2L_0}\frac{\partial}{\partial W}
&\Biggl\{
\frac{1}{\sinh (W-V_0)}
\int_{x-L}^x \rmd z\sinh[W-V(z)]
\nonumber \\
\fl
& \quad \times
\mathcal{D}\int_{z-L}^z \rmd z'
\left(
\rme^{-W}[\hbox{$+$}]_{z'}^z-\rme^{W}[\hbox{$-$}]_{z'}^z
\right)
g(z',W)
\Biggr\}.
\end{eqnarray}
Substituting 
\begin{equation}
\sinh[W-V(z)]=\frac{1}{2}\frac{\partial}{\partial z}
\left(
\rme^{-W}[\hbox{$+$}]_z^x-\rme^{W}[\hbox{$-$}]_z^x
\right),
\end{equation}
and integrating by parts, we obtain
\begin{eqnarray}
\label{1-b.6}
\fl
\int_{x-L}^x \rmd z\sinh[W-V(z)]\int_{z-L}^z \rmd z'
\left(
\rme^{-W}[\hbox{$+$}]_{z'}^z-\rme^{W}[\hbox{$-$}]_{z'}^z
\right)
g(z',W)
\nonumber \\ 
=-\frac{1}{2}\left(\rme^{-W} P- \rme^W M \right)
\int_{x-L}^x\rmd z'
\left(
\rme^{-W}[\hbox{$+$}]_{z'}^x-\rme^{W}[\hbox{$-$}]_{z'}^x
\right)
g(z',W)
\nonumber \\
\qquad
+\frac{1}{2}\int_{x-L}^x \rmd z
\left(
\rme^{-W}[\hbox{$+$}]_z^x-\rme^{W}[\hbox{$-$}]_z^x
\right)
\left(\rme^{-W}P -\rme^W M\right)g(z,W)
\nonumber \\
\qquad 
-\int_{x-L}^x \rmd z
\left(
\rme^{-W}[\hbox{$+$}]_z^x-\rme^{W}[\hbox{$-$}]_z^x
\right)
\sinh[V(x)-W] \int_{z-L}^z \rmd z' \,g(z',W)
\nonumber \\
=0,
\end{eqnarray}
where we have used (\ref{1-5.2}).
In a similar way, using $\rme^{-W}P -\rme^{W}M=2L_0\sinh(V_0-W)$ and $\rme^{-V_0}P -\rme^{V_0}M=0$, we have
\begin{eqnarray}
\label{1-b.7}
\fl
\int_{x-L}^x \rmd z\sinh[W-V(z)]\int_{z-L}^z \rmd z'
\left(
\rme^{-V_0}[\hbox{$+$}]_{z'}^z-\rme^{V_0}[\hbox{$-$}]_{z'}^z
\right)
g(z',V_0)
\nonumber \\
=L_0 \sinh(W-V_0)
\int_{x-L}^x\rmd z'
\left(
\rme^{-V_0}[\hbox{$+$}]_{z'}^x-\rme^{V_0}[\hbox{$-$}]_{z'}^x
\right)
g(z',V_0).
\end{eqnarray}
Equations (\ref{1-b.6}) and (\ref{1-b.7}) give
\begin{eqnarray}
\fl
\frac{1}{\sinh(W-V_0)}
\int_{x-L}^x \rmd z\sinh[W-V(z)]
\,\,\mathcal{D}\int_{z-L}^z \rmd z'
\left(
\rme^{-W}[\hbox{$+$}]_{z'}^z-\rme^{W}[\hbox{$-$}]_{z'}^z
\right)
g(z',W)
\nonumber \\ 
=L_0\int_{x-L}^x\rmd z'
\left(
\rme^{-V_0}[\hbox{$+$}]_{z'}^x-\rme^{V_0}[\hbox{$-$}]_{z'}^x
\right)
g(z',V_0).
\end{eqnarray}
Since this expression is independent of $W$, we can see that the right-hand side of (\ref{1-b.4}) is zero.
Thus, (\ref{1-6.4}) is obtained.


\end{document}